\documentclass{sig-alternate}

\usepackage{fainekos-macros}
\usepackage{amsmath,amssymb,amsfonts,wrapfig,floatflt,yfonts}
\usepackage{mathtools}
\usepackage{colonequals}
\usepackage{enumerate}
\usepackage{graphicx}
\usepackage{subcaption}
\graphicspath{ {figures/} }
\DeclareGraphicsExtensions{.pdf}
\usepackage{multirow}
\RequirePackage{todonotes}
\usepackage{algorithm}
\usepackage{algpseudocode}
\usepackage{textcomp}
\usepackage{color}
\usepackage{url}

\newcommand \CTD {[0,T]}

\newcommand{\teps}{$(\tau,\varepsilon)$}
\newcommand{\tteps}{(\tau,\varepsilon)}
\newcommand{\te}{$(T, J, (\tau,\varepsilon))$}
\newcommand{\tec}{\te-closeness}

\newboolean{MODES}
\setboolean{MODES}{false}

\begin{document}
%

\title{Conformance Testing as Falsification for Cyber-Physical Systems}

\numberofauthors{1} 
\author{
\alignauthor Houssam Abbas, Bardh Hoxha, and Georgios Fainekos\\
\affaddr{CPS Lab, Arizona State University, Tempe, AZ, USA} \\
\email{\{hyabbas, fainekos, bhoxha\}@asu.edu}
\and
\alignauthor Jyotirmoy V. Deshmukh, James Kapinski, and Koichi Ueda,\\
       \affaddr{Toyota Technical Center, Gardena, CA, USA}\\
\email{\{jyotirmoy.deshmukh, jim.kapinski, koichi.ueda\}@tema.toyota.com}
}

\maketitle

\begin{abstract}

In Model-Based Design of Cyber-Physical Systems (CPS), it is often desirable to develop several models of varying fidelity.
Models of different fidelity levels can enable mathematical analysis of the model, control synthesis, faster simulation etc.
Furthermore, when (automatically or manually) transitioning from a model to its implementation on an actual computational platform, then again two different versions of the same system are being developed.
In all previous cases, it is necessary to define a rigorous notion of conformance between different models and between models and their implementations.
This paper argues that conformance should be a measure of distance between systems.
Albeit a range of theoretical distance notions exists, a way to compute such distances for industrial size systems and models has not been proposed yet.
This paper addresses exactly this problem.
A universal notion of conformance as closeness between systems is rigorously defined, and evidence is presented that this implies a number of other application-dependent conformance notions. 
An algorithm for detecting that two systems are not conformant is then proposed, which uses existing proven tools. 
A method is also proposed to measure the degree of conformance between two systems.
The results are demonstrated on a range of models.
\end{abstract}

\section{Introduction}
\label{sec:intro}

\begin{figure}[t]
\centering
\includegraphics[width=7cm]{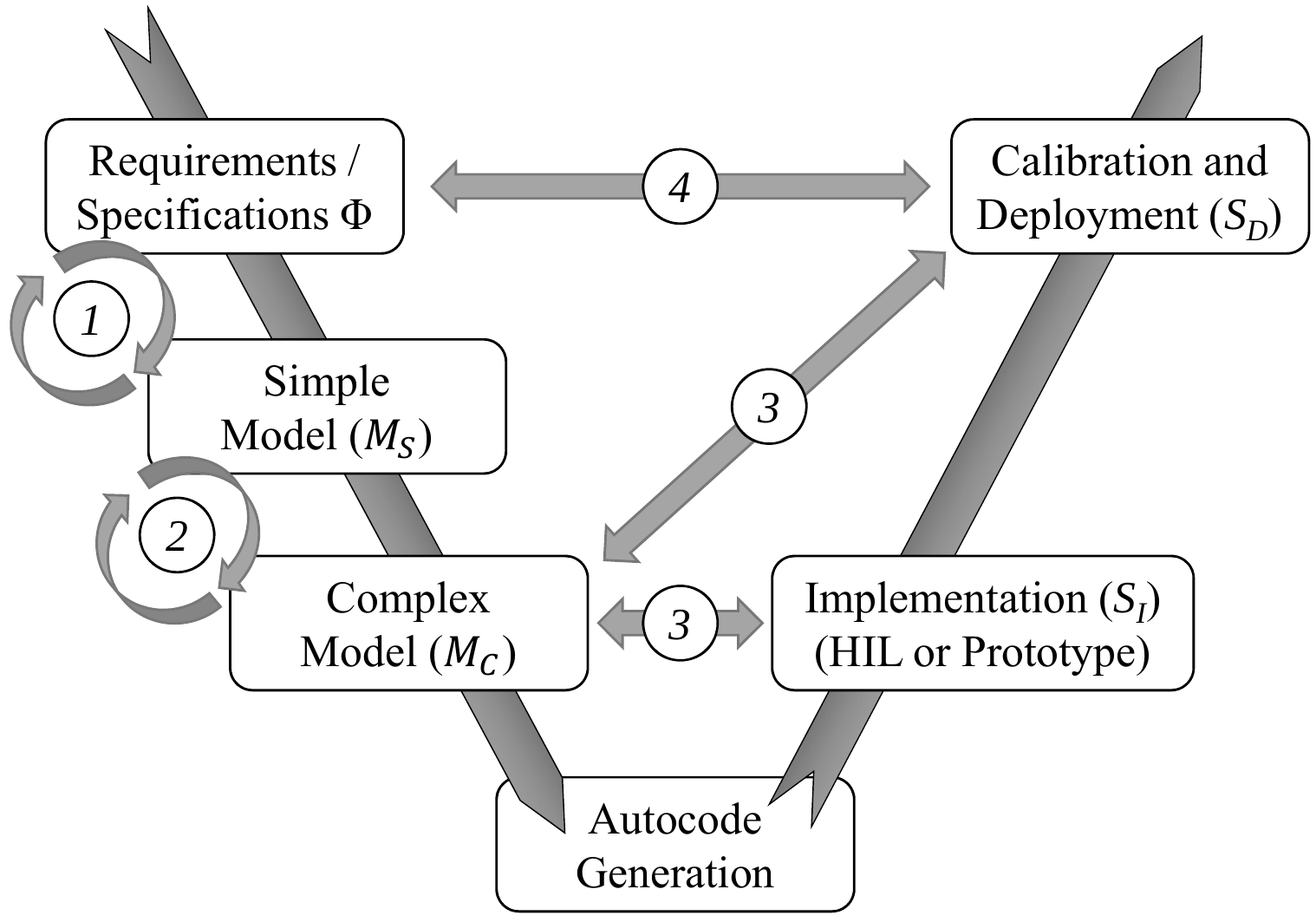}
\caption{Typical V process in MBD. (1) Verifying that the simple model satisfies the functional requirements; (2) Establishing a relationship between the simple and complex model; (3) Verifying conformance of implementation to the model; (4) Verifying that the end product satisfies the functional requirements.}
\vspace{-15pt}
\label{fig:v_process}
\end{figure}

In a typical Model-Based Design (MBD) process for Cyber-Physical Systems (see Fig. \ref{fig:v_process}), a series of models and implementations are iteratively developed such that the end product satisfies a set of functional requirements $\Phi$.
Ideally, the initial (simpler) model $M_S$ developed should have structural properties that make it amenable to formal synthesis and verification methods \cite{Tabuada2009,LeeS11book} (cycle 1 in  Fig. \ref{fig:v_process}) through software tools like \cite{FrehseCAV11,RoyTM11hscc,WongpiromsarnEtAl2011hscc,HuangM2012,Tiwari12cav,PlatzerQ08ijcar}.
Then, the fidelity of the models is increased by modeling more complex physical phenomena ignored initially and by introducing inaccuracies due to the computational platforms such as look-up-tables, time delays, $3^{rd}$ party black-box components, etc.

The development of a higher fidelity model raises the obvious question of what is the relationship between the ``simple" $M_S$ and ``complex" $M_C$ models developed (cycle 2 in  Fig. \ref{fig:v_process}).
If the simpler model developed was a nondeterministic model and the structure of $M_C$ was fully known, then the answer to the question could be established through behavioral inclusions \cite{Tabuada2009}, i.e., is it true that every behavior of $M_C$ can be exhibited by $M_S$, in response to the same stimulus?

However, in practice, non-deterministic models are rarely utilized and supported by industry tools for MBD such as LabView$^{TM}$ or Simulink/Stateflow$^{TM}$.
Instead, a hierarchy of deterministic models is developed each capturing a more accurate representation of the final system, and it is important to know how `close' two successive models are to each other. 
While the higher fidelity model introduces new, more realistic behavior, it should still follow, roughly, the behavior of $M_S$. 
Thus, in lieu of behavioral inclusion, an appropriate notion of \emph{distance} between the models is required, i.e., $\mathbf{dist}( M_C, M_S)$.
This we call conformance between the simple and complex models.
Such distance\footnote{Note we don't use the word `distance' in the mathematical sense.} notions have been developed for various classes of systems \cite{Tabuada2009,Girard08,Antoulas2000,MazziEtAl08cdc,HenzingerMP_FORMATS05} over the years.
Even though works such as \cite{Girard08,MazziEtAl08cdc} treat systems with hybrid dynamics directly, they apply only to certain classes of hybrid systems and, most importantly, they rely on the full knowledge of the mathematical model of both $M_S$ and $M_C$.
For industrial size CPS models, such knowledge is not always available.
Another limitation is that existing distance measures for systems either consider only distances in time, e.g., \cite{HenzingerMP_FORMATS05}, or in space \cite{Girard08,Antoulas2000,MazziEtAl08cdc}. 
For CPS, both are extremely important especially if the end goal is to verify that the deployed system ($S_D$) satisfies formal specifications that involve timing requirements \cite{Koymans90,MalerNickovic04}.

The same observations hold for the important problem of verifying whether a system $S_I$, which is an implementation of a model $M_C$, behaves approximately similar to its model $M_C$ (arrows labeled with 3 in Fig. \ref{fig:v_process}).
Irrespective of whether the automatic code generation process has formal guarantees, rarely does the model $M_C$ capture accurately all physical phenomena.
Thus, the prototype system $S_I$ will be manually modified and calibrated into a final deployment $S_D$.
Then, the deployment $S_D$ should have a bounded, computable distance from the model $M_C$ under an appropriate metric, i.e., $\mathbf{dist}( M_C, S_D) \leq \varepsilon$, and, $S_D$ should satisfy the set of specifications.

In this paper, a framework is provided to address the aforementioned gaps in MBD for CPS, i.e., arrows 2 and 3 in the V process in Fig. \ref{fig:v_process}.
The framework is agnostic about whether the systems studied are both models or a model and its implementation, thus we will generically refer to one system as the Model and to the other as its Implementation.

More specifically, we utilize hybrid distance measures similar to \cite{LygerosJSZS03tac,SanfeliceT10automatica,Caspi2002} in order to define distances between system behaviors.
Given two system behaviors (or trajectories), we compute a $\tteps$ distance between them that captures both their distances in time and in space.
Then, given a bound $(\bar\tau,\bar{\varepsilon})$, we consider the problem of whether the Implementation conforms to its Model with degree $(\bar\tau,\bar{\varepsilon})$.
We pose the aforementioned problem as an optimization problem which we solve using our tool S-Taliro \cite{AnnapureddyLFS11tacas,Fainekos_staliro}.
Our solution is a best effort framework and the guarantees provided are of a probabilistic nature as described for instance in \cite{AbbasF_HybridSA12,AbbasFSIG13tecs,Lecchini10_ConvRates}.

\paragraph{Conformance testing versus specification checking}
\label{sec:conformance vs satisfaction}
One question naturally arises at this point: why not just verify that the Implementation satisfies the same specification that the Model has been verified to satisfy? 
The reasoning behind the question is that if $\Phi$ is all that matters, it should be sufficient that the Implementation also satisfies it.
We may answer this question as follows:
\begin{enumerate}
\item It is not always possible to verify formally that the Implementation satisfies the formal specification: for example, a component purchased from a third party might allow only limited observability and not lend itself to formal methods. 
\vspace{-0.3cm}
\item Parts of the specification are not formally expressed. For example, because the available formal tools can not handle the size of the design (e.g. reachability tools for nonlinear systems). Rather, the specification exists in plain language Test Plan documents \cite{peet_verifPlan04} or implicitly in test suites.\footnote{Note the release of industrial tools that induce requirements from simulation traces, such as \cite{bugscope}, in an effort to formalize requirements currently implicit in tests.}
\vspace{-0.3cm}
\item For a real-life CPS, much of the behavior is {\it de facto} left unspecified because of the complexity. Once triggered, a particular behavior may exhibit unspecified but undesired characteristics, even though it possesses the specified, desired, characteristics (and none of the specified, undesired characteristics).
\end{enumerate}

Therefore, once we have an Implementation, it is not sufficient to check that it too conforms to the specification (if that is even possible). 
It is important to make sure that behaviors exhibited by Model and Implementation are close (in a sense to be defined). This then is conformance testing.
This way, both Model and Implementation display similar unspecified characteristics, and our level of confidence in the Implementation derives from our confidence in the Model.

\subsection{Summary of contributions} 
In the previous sections, we have argued that current ways of thinking about the relation between a Model and its Implementation are not sufficient for the verification of complex CPS. 
In the remainder of this paper, 
\begin{enumerate}
\item We propose a universal definition of conformance between CPSs as a quantifiable closeness measure between the output behaviors of the two systems. 
\item We argue that this universal notion implies most custom conformance notions which depend on the application.
\item We pose conformance testing as a logic property falsification problem. We then apply existing tools to this problem and show that they successsfully find non-conformant behavior.
\item We show that conformance satisfies a monotonicity property which allows us to search efficiently for the best conformance degree between two systems.
\end{enumerate}

\section{Problem Formulation}
\label{sec:problem}

In Section \ref{sec:intro}, it was argued that the verification of a CPS implementation in an MBD process requires conformance testing. 
The latter was described as checking that Model and Implementation display `similar' behaviors, where `similar' will be made precise.
Because the objective is to detect bugs caused by the implementation process, Model and Implementation should be tested with the same inputs, and starting from the same initial conditions.
Our high level goal is then to determine whether there exists a pair of (initial conditions, input signal) that causes the Model and its Implementation to produce significantly different outputs; and if such a pair exists, to find it and present it to the user as a debug guide.

To make this goal precise, this section starts by presenting the class of systems that we study.
This class is illustrated with a running example of a fuel control system for an automotive application.
Then, the conformance testing problem is formally stated as a search problem over the set of initial conditions and input signals.
Finally, the constraints under which we seek to solve this problem are presented.
This lays the groundwork for Section \ref{sec:defining conformance}, where we will mathematically define what it means for two CPSs to be conformant.

\emph{Notation}. 
Given two sets $A$ and $B$, $B^A$ denotes the set of all functions from $A$ to $B$.
That is, for any $f \in B^A$ we have $f : A \rightarrow B$.
Given a cartesian set product $A\times B$, $\proj_A$ is the projection onto $A$, i.e. for all $(a,b)\in A\times B, \proj_A((a,b))=a$.

\subsection{System model and running example}
\label{sec:system model}

At its most general, a CPS $\Sys$ may be thought of as an input-output map. 
Specifically, let $N = \{1,2,\ldots,|N|\} \subset \Ne$ be a finite set of integers,
$T > 0$ be a positive real,
$\hsSet_0 \subset \Re^{n_h}$ be a set of {\it initial operating conditions} of the system, 
$\inpSet \subset \Re^{n_u}$ be a compact set of \emph{input values},
and let $\hoSet \subset \Re^{n_y}$ be a set of \emph{output values}.

\begin{defn}
\label{def:tss}
A \textbf{real-timed state sequence} (real-TSS) is a pair $(\hotraj,\htstmp)$ where $\hotraj \in \hoSet^{|N|}$ and $\htstmp \in \CTD^{|N|}$. \\
A \textbf{hybrid-timed state sequence} (hybrid-TSS) is a pair $(\hotraj, \htstmp)$ where $\hotraj \in \hoSet^{|N|}$ and $\htstmp \in (\CTD \times \Ne)^{|N|}$.\\
\end{defn}
When a statement applies to both real-timed and hybrid-timed state sequences, we will simply say `timed state sequence' (TSS).
A TSS can be the result of a sampling process or a numerical integration. Then the vector of `timestamps' $\htstmp$ represents the sequence of sampling times, or times at which a numerical solution is computed.
A timed state sequence will also be referred to as a \emph{signal}, and a $\hoSet$-valued timed state sequence will also be referred to as a \emph{trajectory}. The latter is standard dynamical systems theory terminology.
Note that a real-TSS may be viewed as a special hybrid-TSS $(\hotraj,\htstmp)$ such that $\htstmp \in (\CTD\times \{1\})^{|N|}$.

A CPS is modeled as a map between initial conditions $h_0 \in \hsSet_0$ and \emph{input} timed state sequences $(\inpPt,\tstmp_{\inpSig}) \in \inpSet^{|N|} \times \Te^{|N|} \defeq \inpSpace$ 
to \emph{output} timed state sequences $(\hotraj,\tstmp_{\hotraj}) \in \hoSet^{|N|} \times \Te^{|N|}$,
where $\Te$ is either $\CTD$ (for real-timed) or $\CTD \times \Ne$ (for hybrid-timed).
Note that input and output signals must either both be real-timed, or both be hybrid-timed.
We model discrete states as integers, so $\hsSet, \inpSet$ and $\hoSet$ could be hybrid spaces of the form $X \times L$ with $X \subset \Re^n$ and $L$ finite.
The system $\Sys$ can then be viewed as a map:
\begin{equation}
\Sys: (h_0, \inpSig) \in \hsSet_0 \times \inpSpace \mapsto (\hotraj,\htstmp) \in \hoSet^{|N|} \times \Te^{|N|}
\label{eq:sys}
\end{equation}

We impose the following restrictions on the systems that we consider:
\begin{enumerate}
\vspace{-0.1cm}
\item The output space $\hoSet$ must be equipped with a generalized metric $\genMet$. See \cite{AbbasFSIG13tecs} for implications.
\label{ass:2}
\vspace{-0.2cm}
\item For every initial condition $\hsPt_0 \in \hsSet_0$ and input signal $\inpSig \in \inpSpace$, the system $\Sys$ produces an output signal.
This is imposed to avoid modeling issues where the Model's and/or Implementation's equations have no solutions.
\label{ass:3}
\end{enumerate}
\vspace{-0.2cm}
Further details on the necessity and implications of the aforementioned assumptions can be found in \cite{AbbasFSIG13tecs}.


As it is standard in systems theory, the system's output can be expressed as a function of its internal state $\hsPt \in H \supset \hsSet_0 $:
\[ \hsPt \in \hsSet \mapsto \hoPt = g(\hsPt) \in \hoSet\]
Here, $\hsSet$ is the state-space of the system. We do not always assume that the internal state is observable.
Given a real-timed state sequence $(\hotraj,\tstmp)$, its $i^{th}$ element is denoted $(\hotraj,\tstmp)_i = (\hotraj_i,\tstmp_i) \in \hoSet \times \Te$.
Similarly, given a hybrid-timed state sequence $(\hotraj,\htstmp)$, the $i^{th}$ element $(\hotraj,\htstmp)_i$ is denoted $(\hotraj_i,\htstmp_i)$, with $\htstmp_i = (t,j) \in \CTD \times \Ne$.

\begin{exmp}
\label{ex:AFC}
We consider a fuel control (FC) system for an automotive application.
 Environmental concerns and government legislation require that the
 fuel economy be maximized and the rate of emissions (e.g.,
 hydrocarbons, carbon monoxide, and nitrogen oxides) be minimized.
 Control of automobile engine air-to-fuel (A/F) ratio is crucial to
 optimize fuel economy and to minimize emissions. Ideal A/F levels are
 given by the {\em stoichiometric} value, which is the optimal A/F
 ratio to minimize both fuel consumption and emission of pollutants.
 The purpose of the FC system is to maintain the ratio of air-to-fuel
 (A/F) within a given range of the stoichiometric value.

 The scenario that we model involves an engine connected to a
 dynamometer, which is a device that can control the speed of the
 engine and measure the output torque. For our experiment, the
 dynamometer maintains the engine at a constant rotational velocity, as
 the engine is tested. There is only one input to the model: the
 throttle position command from the driver.

 The conformance testing scenario for this example is unique, in that
 the Model was derived from the Implementation, for reasons on which we
 will now elaborate. The Implementation was derived from a textbook
 model of an engine control system \cite{Guz2010}, and contains
 implementation details such as look-up-tables (LUTs). The Model was then
 abstracted from this Implementation for the purposes of formal
 analysis \cite{Jin14}.

 Despite the counter-intuitive relationship between the Model and
 Implementation for this case, the conformance task remains: to verify
 that these two versions satisfy some similarity criterion.
\exmend
\end{exmp}

The discussion and results in this paper apply to this input-output map model of a CPS.
To define some of the conformance notions in this paper, it will be useful to sometimes work with the more specialized \emph{hybrid automaton} model of a CPS~\cite{LygerosJSZS03tac}: broadly speaking, a hybrid automaton has countably many modes $\{\mode_1,\mode_2,\ldots\} \defeq \modeSet$, 
with possibly different dynamics $F_{\mode}$ active in each mode: $\dot{\hsPt} = F_\mode(\hsPt, \inpPt)$. 
The automaton switches (or `jumps') between modes whenever the internal state $\hsPt$ enters specific subsets $G \subset \hsSet$ of the state space, called \emph{switching guards}. 
In general, a switching guard might depend on time and on the current state; different jumps $\mode \rightarrow \mode'$  will have different guards: $G = G(e, t,\hsPt)$, $e = (\mode,\mode')$.
Finally, when the system switches modes, the internal state might be reset to a switch-specific value: $\hsPt^+ = Re(\hsPt,e)$ if $\hsPt \in G(e,t,\hsPt)$.
If we explicitly model the system mode as part of the internal state $\hsPt=  (\stPt,\mode) \in \stSet \times \modeSet$, we may write the automaton's equations as \cite{Sanfelice11_interconnections}
\begin{equation}
\label{eq:HA}
\Sys \left\{ \begin{array}{lll}
(\dot{\stPt},\dot{\mode}) &= (F_{\mode}(\stPt, \inpPt), 0)  &\quad (\stPt, \inpPt) \in C \times \inpSet \\
(\stPt^+, \mode^+)        &= Re(e,\stPt)         &\quad (\stPt, \inpPt) \in D \times \inpSet \\
\hoPt 									  &= g(\hsPt)
\end{array}
\right.
\end{equation}
where $C \subset \stSet$ is the `flow set' of continuous evolution, 
and $D$
is the jump set, which equals the union of all guard sets.
Apart from the requirement that the dynamics have at least one solution for every $(\hsPt_0,\inpSig)$, they are arbitrary.

\begin{rem}
The notion of a system mode applies to the general input-output model of a system, so in what follows we will often be referring to the `mode' of the CPS without necessarily requiring that it be modeled as a hybrid automaton.
For example, a powertrain Implementation might be outputting the current gear, or the mode of operation e.g. Economy vs. Sport.
\end{rem}

The trajectories (or `solutions') of purely continuous dynamical systems (with only one mode) are parameterized by the time variable $t$, and those of purely discrete dynamical systems (with no continuous evolutions) are parametrized by the number of discrete jumps $j$. 
Following Goebel and Teel \cite{GoebelT06_SolnsHybInclusions}, the trajectories to hybrid automata are parametrized by both $t$ and $j$, to reflect that both evolution mechanisms are present.
So we write $\hstraj(t,j)$ for the state and $\hotraj(t,j)$ for the output of the automaton at time $t$ and after $j$ jumps, or mode switches.
Because jumps take 0 time, it is possible to have the automaton go through several states in 0 time: $\hstraj(t,j) \rightarrow \hstraj(t,j+1) \rightarrow \hstraj(t,j+2) \ldots$. This can't happen in a physical Implementation, but it may be allowed in the Model.
We refer the reader to \cite{GoebelT06_SolnsHybInclusions} for exact definitions of discrete and hybrid time domains, arcs and trajectories.

We now introduce the behavior of a system which is applicable to both input-output maps and hybrid automata.
\ifthenelse{\boolean{MODES}}
{
\begin{defn}[Behavior]\label{def:behavior}
Take a system $\Sys$, an initial point $h_0 \in \hsSet$ and input signal $\inpSig$.
The behavior of the CPS $\Sys$ from $h_0$ and $\inpSig$, denoted $\behavior_\Sys(h_0,\inpSig)$,
consists of 
\begin{itemize}
	\item $h_0 = (\stPt_0, \mode_0)$
	\item the output trajectory $\hotraj_\Sys(h_0,\inpSig)$ generated by $\Sys$ in response to $(h_0,\inpSig)$
	\item the mode trajectory $\modetraj(h_0,\inpSig)$ 
	\item the transition times $(t_i)_{i\geq 0}$, $t_0 = 0$, at which the system switches modes
\end{itemize}
The behavior of $\Hc$ is then
\begin{equation*}
\label{eq:behavior}
\behavior_{\Hc} = \{(h,\hotraj_\Sys(h,\inpSig), \modetraj(h,\inpSig), (t_i)_i) \st h \in \hsSet_0, \inpSig \in \inpSpace \}
\end{equation*}
\end{defn}
}
{
\begin{defn}[Behavior]\label{def:behavior}
Take a system $\Sys$, an initial point $h_0 \in \hsSet$ and input signal $\inpSig$.
The behavior of the CPS $\Sys$ from $h_0$ and $\inpSig$, denoted $\behavior_\Sys(h_0,\inpSig)$,
consists of 
\begin{itemize}
	\item $h_0 = (\stPt_0, \mode_0)$
	\item the output trajectory $\hotraj_\Sys(h_0,\inpSig)$ generated by $\Sys$ in response to $(h_0,\inpSig)$
\end{itemize}
The behavior of $\Hc$ is then
\begin{equation*}
\label{eq:behavior}
\behavior_{\Hc} = \{(h,\hotraj_\Sys(h,\inpSig)) \st h \in \hsSet_0, \inpSig \in \inpSpace \}
\end{equation*}

\end{defn}
}

\begin{exmp}[Example \ref{ex:AFC} Continued]
For the FC, the outputs consist of the normalized air-to-fuel ratio $\lambda$ and the fuel commanded into the Cylinder-and-Exhaust. Thus $\hoSet = \Re_+^2$.
The presence of a switch in the Throttle block, and an LUT in the Cylinder-and-Exhaust block, induces 8 modes so $L = \{1,\ldots,8\}$.
The outputs of the FC are sampled at a fixed rate.
The output signals can be modeled as real-TSS. If we could observe the mode changes during a simulation, then we can use a counter $j$ to count the mode switches, or `jumps', and model the output as a hybrid-TSS.
E.g. the following sequence of sampled $(\lambda,\mbox{mode})$
\[(1,\mode_1), (0.99,\mode_1), (0.98,\mode_2), (0.87,\mode_2), (0.88,\mode_1)\]
is interpreted as the following hybrid-TSS
\begin{align*}
(\hotraj, \htstmp)  &= ((1,0.99,0.98,0.87,0.88), \\
 &((0,0), (T/|N|,0),(2T/|N|,1),(3T/|N|,1),(3T/|N|,2))
\end{align*}
Note that $j$ counts the jumps so far, but does not indicate what mode the system is in.
\end{exmp}

\subsection{Conformance testing}
\label{sec:conformance testing}
Based on the preceding discussion, we adopt the premise that conformance is a relation of similarity between the behaviors of two systems when subjected to the same stimulus. 
The behavior is defined in Def.~\ref{def:behavior}. So we may speak of conformant (i.e., similar) behaviors, or conformant output trajectories. 
Conformance testing can then be formulated as a search problem: find a pair of trajectories, generated by the two CPSs in response to the same initial condition and input, that are not conformant. 
The search for a non-conformant pair of trajectories is called \emph{falsification}.

\begin{prob}[Conformance testing]
Let $\Sys_M$ and $\Sys_I$ be a Model and Implementation of a CPS, respectively.
Find a pair $\theta = (\hsPt, \inpSig) \in \hsSet \times \inpSpace$ such that $\hotraj_{\Sys_M}(\theta)$ and $\hotraj_{\Sys_I}(\theta)$ are non-conformant. 
\end{prob}

Because Implementations typically have limited observability, we assume testing happens under the following restriction:
\begin{ass}[Black box testing]
\label{ass:bbox}
The behaviors of Model and Implementation are observable: i.e. it is always possible, for either system, to obtain an element of the behavior by executing the system.
Only the behavior of the Implementation is observable - i.e. we know nothing else about it.
\end{ass}

\ifthenelse{\boolean{MODES}}
{During the debug and development cycles, an Implementation can be instrumented to provide additional visibility such that Assumption \ref{ass:bbox} is satisfied; specifically, the mode sequence can be observed.}
{In particular, the sequence of modes that Model and Implementation go through can be an important variable to track to decide whether the two systems conform.}
As an example, a silicon microchip has `scan chains', which are chains of buffers that pass to the outside world the values of internal registers. These are only used during testing, and are burnt before customer delivery.
In control systems, mode estimation \cite{Bako_ModeEstimation13} could be used when applicable.
While we leave it possible that more is known about the Model, we won't need to know more to apply the methods of this paper.
More knowledge of the Model will make applicable grey box testing methods such as \cite{AbbasF_NonlinearDescent13,AbbasATVA11_LinFalsification}.

\section{Defining conformance}
\label{sec:defining conformance}
In general, \emph{conformance is an application-dependent notion} to help determine that the implementation process does not use components or methods that alter the functionality (or safety or performance) of the final product in any significant manner. 
What `significant' means will, naturally, depend on the application.
This makes conformance testing itself application-dependent.
Our first contribution is made in this section: we present a notion of distance between output trajectories, called \tec, and argue that this is an appropriate \emph{universal} notion of conformance; that is, it is generally applicable regardless of the underlying application.
The price we pay for this universality is that this notion is stronger than the application-dependent ones: two systems may not be conformant according to \tec, but they may be conformant according to a weaker custom notion which is sufficient for the task at hand.
In the second part of this section, we give real-life examples where the application-dependent conformance turns out to be implied by \tec. 

Thus we may develop a general theory of conformance based on \tec, and `generic' algorithms that decide conformance which do not depend on the application. This is advantageous for two reasons: one of the challenges today for testing of hybrid systems (and CPS in general) is to define conformance in a rigorous manner, and \tec\; provides an answer. Secondly, generic conformance tools can be used early in the design cycle, before the instrumentation is all there for a deeper analysis of the difference between Model and Implementation.
Moreover, a feature of the universal notion is that it uses only the outputs of the system (and possibly the mode sequence if available). Thus, the analysis and methods herein are applicable to potentially complicated systems with very general system models, including the input-output map model in Section \ref{sec:system model}.

\subsection{A universal conformance notion}
\label{sec:hyb simulation}
The proposed universal notion of conformance is \tec.
\tec~expresses proximity between the outputs, their time sequences (real-TSS and hybrid-TSS), and their modes if applicable.
It is derived from~\cite{GoebelT06_SolnsHybInclusions}. 
\begin{defn}[\tec]
\label{def:tecloseness}
Take a test duration $T > 0$, a maximum number of jumps $J \in \Ne$, and parameters $\tau, \varepsilon >0$. 
Two timed state sequences, or \emph{trajectories}, $(\hotraj,\htstmp)$ and $(\hotraj',\htstmp')$ are \te-close  if \\
(a) for all $i \in N$ such that $\htstmp_i = (t,j)$ satisfies $t \leq T, j \leq J$, there exists $k \in N$ such that $\htstmp'_k = (s,j)$, $|t-s|<\tau$, and 
\[\|\hotraj_i - \hotraj'_k\| < \varepsilon \] \\
(b) for all $i \in N$ such that $\htstmp'_i = (t,j)$ satisfies $t \leq T, j \leq J$, there exists $k \in N$ such that $\htstmp_k = (s,j)$, $|t-s|<\tau$, and 
\[\|\hotraj'_i - \hotraj_k\| < \varepsilon \] \\
We will also say that $\hotraj_1$ and $\hotraj_2$ are \textbf{conformant} with degree \te.
\end{defn}
When $T$ and $J$ are clear from the context, we simply say \teps-close. 
Because a real-TSS is a special case of a hybrid-TSS, the above definition applies to both.

\tec\; may be tought of as giving a proximity measure between the two hybrid arcs, both in time and space. The definition says that within any time window of size $2\tau$, there must be a time when the trajectories are within $\varepsilon$ or less of each other.
Allowing some `wiggle room' in both time and space is important for conformance testing: when implementing a Model, there are inevitable errors. These are due to differences in computation precision, clock drift in the implementation, the use of inexpensive components, unmodeled environmental conditions, etc, leading to the Implementation's output to differ in value from the Model's output, and to have different timing characteristics.
Thus \tec\; captures nicely the intuitive notion that `the outputs should still look alike'. 
Our definition of \tec\; differs slightly from the original definition in ~\cite{GoebelT06_SolnsHybInclusions} in that we use two `precision' parameters $\tau$ and $\varepsilon$ instead of one. 
In practice, using only one precision parameter is too restrictive, since the outputs can have a different order of magnitude from the time variable.
It can be verified that the \textsc{hioco} relation of Van Osch \cite{Osch_IOCO06} is an exact version of \tec\; ($\tau = \varepsilon = 0)$, with the role of inputs and outputs explicitly differentiated. 

\begin{rem}
\label{rem:teps no j}
If it is not possible to observe the number of jumps $j$, then we simplify the above definition by assuming that $j$ is always equal to 1. 
In other words, we interpret the definition over real-TSS and assume the system only has one mode.
\end{rem}

\begin{defn}
\label{def:tecsystem}
Take a test duration $T > 0$, a maximum number of jumps $J \in \Ne$, and parameters $\tau, \varepsilon >0$. 
Two CPSs $\Sys_M$ and $\Sys_I$ are said to be \te-close if for any initial condition $h_0 \in \hsSet_0$ and input signal $\inpSig \in \inpSpace$, the trajectories $\hotraj_{\Sys_M}(h_0,\inpSig)$ and $\hotraj_{\Sys_I}(h_0,\inpSig)$ are \te-close. 
The two systems are also said to be conformant with degree \te.
\end{defn}

\begin{rem}
A Model and Implementation generally won't have the same state-space, and so won't accept the same initial conditions. So when we provide the same initial condition $h_0$ to both, one of them might use a projection of $h_0$ or a more general mapping $f(h_0)$ to obtain its appropriate initial conditions.
\end{rem}

From a conformance perspective, it is preferable to have a smaller $\varepsilon$ and a smaller $\tau$.
We use this to define a partial order on the $\tteps$ pairs.
\begin{defn}
\label{def:teorder}
The partial order relation $\preceq$ over $\tteps$ pairs is given by
$\tteps \preceq (\tau',\varepsilon')$ if and only if $\tau \leq \tau'$ and $\varepsilon \leq \varepsilon'$.
%
The inequality is strict if and only if at least one of the component-wise inequalities is strict.\\
%
\end{defn}


\begin{rem}
\label{rem:monotonous}
\tec~has the valuable advantage of being monotonic: if two trajectories are \te-close, then they are $(T,J,(\tau',\varepsilon'))$-close for any $(\tau',\varepsilon') \succeq \tteps$.
This allows us to use a simple binary search for a smallest $\tteps$ pair such that the trajectories, and the systems, are \te-close.
We make use of this property in the experiments.
\end{rem}

\subsection{Examples}
\label{sec:appconf}

We conclude this section with examples where application-specific notions of conformance are implied by \tec.
Thus if we find trajectory pairs $(\eta_{\Sys_M},\eta_{\Sys_I})$ that violate the latter, they automatically violate the former. 
\begin{exmp}[Example \ref{ex:AFC} continued]
\label{ex:afccustom}
Because the look-up-tables (LUTs) in the Implementation $\Sys_I$ are replaced by polynomials in the Model $\Sys_M$, some error is expected between the outputs of the two systems. 
The designer hopes, however, that the error at the output of the Implementation, is in the same order of magnitude as the error between the outputs of the LUTs and the outputs of the corresponding polynomials. 
If not, then more entries are needed in the LUT.
Moreover because LUT look-ups are typically faster than polynomial computations, some delay between the two outputs is expected to be observed. The designer has a pre-specified maximum acceptable delay.
 In this case, conformance imposes upper bounds on the spatial and temporal differences between the outputs of Model and Implemenation.
\end{exmp}


Conformance testing is applicable to application domain areas other than the automotive industry. E.g. in the microchip design cycle, as shown in the following example. 

\begin{exmp}[State retention]
\label{ex:stateretention}
$\Sys_M$ is an RTL description of an electrical circuit, and $\Sys_I$ is equal to $\Sys_M$ with power gating and state retention added to some of its subsystems. 
With state retention, the contents of certain critical memory elements
of the power-gated subsystem are retained in `shadow' registers prior to power-down, and restored after power-up. 
This creates a temporary difference between the state of the non-state retained circuit ($\Sys_M$) and the state-retained circuit ($\Sys_I$). This difference lasts until the reset sequence is completed. 
	Thus in this case, conformance means that a temporary difference in modes between the two systems is allowed, but they must re-converge after a pre-defined amount of time.  
\end{exmp}

\section{Solution approach}
\label{sec:solution}

\newcommand \hsPtp {\hsPt_{||}}
\newcommand \hoPtp {\hoPt_{||}}
\newcommand \hotrajp {\hotraj_{||}}
\newcommand \htstmpp {\htstmp_{||}}
\newcommand \Sysp {\Sys_{||}}

In this section, we present a general method for determining whether two systems are conformant or not.
We also provide a way to quantify the degree of conformance between them.

\subsection{Conformance as falsification}
\begin{figure}
\centering
\includegraphics[width=\columnwidth]{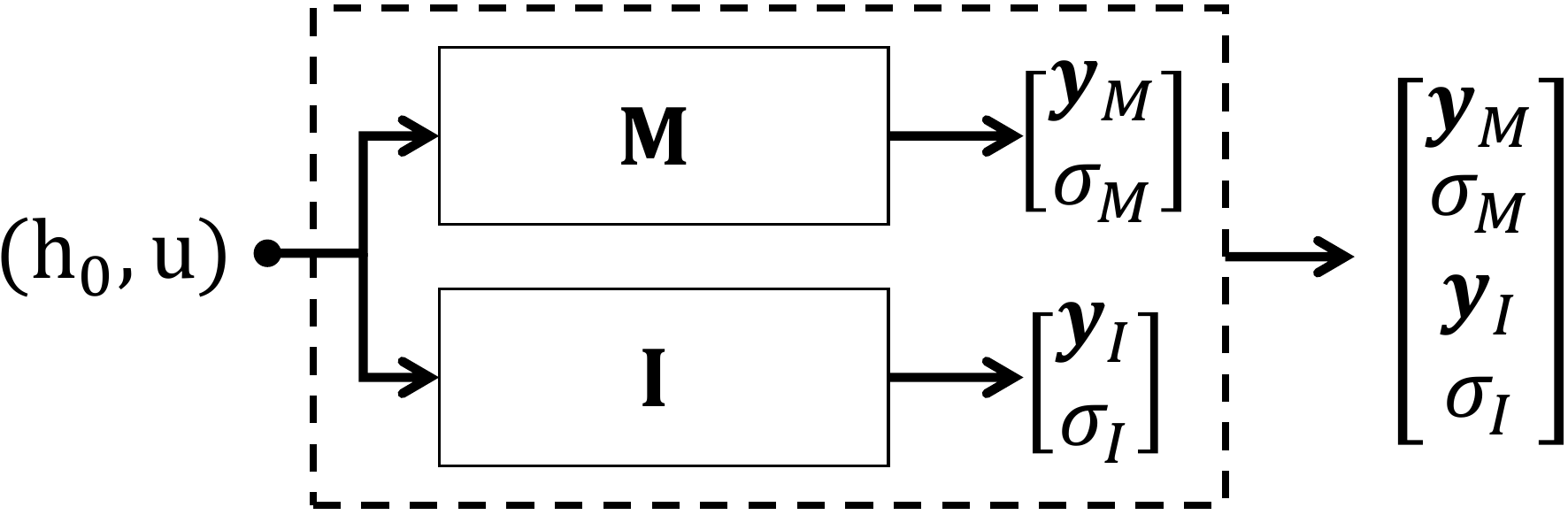}
\caption{Parallel interconnection of Model and Implementation. }
\label{fig:parallel_system}
\vspace{-0.2cm}
\end{figure}

Our approach is based on the observation that \tec~can be expressed as a formal logical property defined over the output timed state sequences of the parallel interconnection of systems $\Sys_M$ and $\Sys_I$. See Fig. \ref{fig:parallel_system}. 
A TSS of the interconnection system $\Sysp$ is just the concatenation of the TSS of the component systems: $(\hotrajp, \htstmpp) = ((\hotraj_M, \htstmp_M), (\hotraj_I, \htstmp_I))$.
If we can find a (parallel) TSS (or `trajectory') $(\hotrajp, \htstmpp)$ which falsifies the \tec~property, then by definition, the component trajectories are non-conformant, and by extension, the systems $\Sys_M$ and $\Sys_I$ are non-conformant.
In what follows, we will use the terms `falsifying trajectory pairs' and `non-conformant trajectory pairs' interchangeably.

The logic we use to express \tec~is Metric Temporal Logic (MTL) \cite{Koymans90} (see Appendix for a review of MTL).
We first present the following construction for real-TSS, then generalize it to hybrid-TSS.

\underline{Real-TSS:} Fix $\tau> 0$, $\varepsilon > 0$.
Our goal is to express \teps-closeness as an MTL formula.
Let $(\hotraj_M, \tstmp_M)$ and $(\hotraj_I, \tstmp_I)$ be the outputs of Model and Implementation CPSs, respectively, in response to the same initial conditions and input signal.
Because \teps-closeness requires comparing the current value of $\hotraj_M$ to current, past and future values of $\hotraj_I$ (over a window of width $2\tau$), we will create shifted versions of $\hotraj_I$. 
Given the symmetry of \teps-closeness, we will also define shifted versions of $\hotraj_M$.
The amount of the shift will depend on $\tau$: how many samples of $\hotraj_I$ ($\hotraj_M$) fit within a window of width $2\tau$?

The shifted versions are now defined.
Recall that $\hotraj_{M,i}$ is the $i^{th}$ sample in the TSS $\hotraj_M$, and similarly for $\hotraj_{I,i}$. 
Consider the Model's output: for each $i \in N$, compute the largest $k \geq i$ such that $|\tstmp_{M,i} - \tstmp_{M,k}| < \tau$. Define $m(\tau,i) = k-i$: this is the number of samples in the largest window of duration less than $\tau$ starting at $\tstmp_i$.
Similarly, we compute $n(\tau,i)$ for the Implementation TSS for every $i \in N$.
The numbers $n$ and $m$ could in general vary with $i$ due to an adaptive sampling period.
Define $m(\tau) = \min\{m(\tau,i), i \in N\}$. $m(\tau)$ is the smallest number of samples in a window of size less than $\tau$ anywhere in $(\hotraj_M, \tstmp_M)$.
Assuming that $|\tstmp_M| > 1$, it comes that $m(\tau) < \infty$.\footnote{The case where $m(\tau) = \infty$ occurs when the Model trajectory $\hotraj_M$ is Zeno: i.e., when it contains an infinite number of samples without advancing time. This can result from a modeling artifact~\cite{Lygeros99_Zeno}. The condition $|\tstmp_M| > 1$ effectively says we have at least two different timesteps, and so the trajectory is not initially Zeno.}
 This constitues the size of the shift (forward and backward) to apply to $(\hotraj_M,\tstmp_M)$.
Similarly, define $n(\tau)$ for the Implementation. 
We may now define shifted versions of the output trajectories via the discrete shift operator: 
for $k \in \Ze$, $\Shift_k (\hotraj_M,\tstmp_M) \defeq (\Shift_k \hotraj_M, \Shift_k \tstmp_M)$, with
\begin{eqnarray}
\Shift_k\hotraj &=& (\hotraj_{k+1},\hotraj_{k+2}, \ldots,\hotraj_{|N|},\underbrace{\hotraj_{|N|},\ldots,\hotraj_{|N|} }_{\text{k times}})		        \label{eq:sky1} \\
\Shift_k\tstmp  &=& (\tstmp_{k+1} ,\tstmp_{k+2} , \ldots,\tstmp_{|N|} ,             \tstmp_{|N|}+\frac{T}{|N|},\ldots,\tstmp_{|N|}+\frac{kT}{|N|} ) \nonumber
\end{eqnarray}
when $k >0$, and 
\begin{eqnarray}
\Shift_k\hotraj &=& (\underbrace{\hotraj_{1},\ldots,\hotraj_{1} }_{\text{k times}}, \hotraj_{1},\hotraj_{k+2},\ldots,\hotraj_{|N|-k})    \label{eq:sky2}\\
\Shift_k\tstmp  &=& \tstmp \nonumber
\end{eqnarray}
when $k < 0$.
Note that the filler values at both ends of the shifted sequences $\Shift_k \hotraj$ \eqref{eq:sky1},\eqref{eq:sky2} are obtained by constant interpolation.

Recall Def.~\ref{def:tecloseness}(a). This condition can be captured by saying that at all $i$, there exists a $k \in \{-n(\tau),\ldots,n(\tau)\}$ such that $\|\hotraj_{M,i} - (\Shift_k \hotraj_I)_i\| < \varepsilon$. 
Analogously for Def.~\ref{def:tecloseness}(b).

Now \tec~may be expressed as the following MTL formula $\formula_{\tteps}$ 
($\lor$ is the logical OR operator, $\land$ is the logical AND operator, and $\always_\Ic$ is the temporal `Always over the time interval $\Ic$' operator - see Appendix)
\begin{eqnarray}
\label{eq:phitaueps}
p_1\tteps & = & \bigvee_{k=-n(\tau)}^{n(\tau)} \|\hotraj_{M,i} - (\Shift_k \hotraj_I)_i\| < \varepsilon \label{eq:p1} \\
p_2\tteps & = & \bigvee_{k=-m(\tau)}^{m(\tau)} \|\hotraj_{I,i} - (\Shift_k \hotraj_M)_i\| < \varepsilon \label{eq:p2} \\
\formula_{\tteps} & \defeq & \always_{[0,T]} (p_1\tteps \land p_2\tteps ) \label{eq:phi}
\end{eqnarray}

Because \te-closeness only requires that the two signals be within $\varepsilon$ of each other at least once in a window of size $2\tau$, $p_1$ and $p_2$ use disjunction: it is sufficient for one shifted comparison to be less than $\varepsilon$.

\underline{Hybrid-TSS:} To define the MTL formula over hybrid-TSS, we must break up each trajectory into segments, such that there are no jumps within a segment. 
Specifically, consider the hybrid-TSS $(\hotraj, \htstmp)$, with 
\[\htstmp = ((t_1,j_1), (t_2,j_2), \ldots,(t_{|N|},j_{|N|}))\] 
Assume that there are only $G$ unique values of $j$ that appear in $\htstmp$, corresponding to $G-1$ jumps. We divide the hybrid-TSS into $G$ segments $g_1,\ldots,g_G$, such that $j$ is constant over a segment. Each segment can be viewed as a real-TSS.
If we apply this procedure to $(\hotraj_M, \htstmp_M)$ and $(\hotraj_I, \htstmp_I)$, we get $G_M$ Model segments $\{g^M_i\}_{i=1}^{G_M}$ and $G_I$ Implementation segments $\{g^I_i\}_{i=1}^{G_I}$. Let $G = \max\{G_M,G_I\}$.
We can now apply the above procedure to every pair $(g^M_i,g^I_i)$, with the important difference that the shifted sequences $\Shift_k \hotraj$ \eqref{eq:sky1},\eqref{eq:sky2} are filled with an arbitrarily large value (or $+\infty$), and not by constant interpolation. This is to reflect that a comparison past the jump point is not valid.
This results in $G$ formulae $\formula_{\tteps}^{i}$ obtained via \eqref{eq:phi}. The complete formula can now be written
\begin{equation}
\label{eq:phitauepsHtss}
\Phi_{\tteps} = \bigwedge_i \formula_{\tteps}^{i}
\end{equation}

Note there are other ways of defining the MTL formula for hybrid-TSS that directly incorporate the jump counter in the formula. Comparing these different methods is outside the scope of this paper.
Unless otherwise indicated, all the discussion that follows applies equally to the formula obtained via \eqref{eq:phi} (for real-TSS) or \eqref{eq:phitauepsHtss} (for hybrid-TSS). 

We can now use existing tools, like {\staliro} \cite{AnnapureddyLFS11tacas,Fainekos_staliro}, to find a pair of trajectories (equivalently, a trajectory of the parallel interconnection) which falsify $\formula_{\tteps}$. 
{\staliro} uses, among others, Simulated Annealing (SA) to find falsifying trajectories.
If such a trajectory is not found, convergence properties of SA imply that with probability approaching 1, the property is satisfied by the systems; equivalently, that the two systems are indeed conformant.

We should stress at this point that the proposed method is not specific to \tec. It is more widely applicable to any application-dependent conformance notion that can be expressed as an MTL formula, including those from the examples in Section \ref{sec:defining conformance}.
For example, for the case when mode sequences are allowed to diverge for at most a pre-defined duration $D$ (Example \ref{ex:stateretention}), the conformance relation is expressed as:
``For every initial condition $h_0 \in \hsSet_0$ and every input signal $\inpSig \in \inpSpace$, whenever the two systems are in different modes, they will be back in the same mode within $D$ sec''.
This can now be written as the MTL formula:
\begin{equation}
\label{eq:npwc}
\formula_{PWC} \defeq \always_{[0,T-D]} (\ell(t) \ne \ell'(t) \Rightarrow \eventually_{[0,D]} \ell(t) = \ell'(t)) \\
\end{equation}

We conclude this section with a word on how to practically falsify $\formula_{\tteps}$ (or any of the other application-dependent notions).
A method that has proved efficient is to minimize the \emph{robustness} of the trajectories w.r.t the MTL property. 
In this work, we use spatial robustness \cite{AbbasFSIG13tecs,FainekosP09tcs} and time robustness \cite{DonzeM_SignalTL10}.
Spatial robustness measures how far \emph{in the output space} a given trajectory is from the nearest trajectory with opposite truth value for $\formula$.\footnote{If the mode is observable, spatial robustness also computes the (quasi-) distance between the modes of the two trajectories~\cite{AbbasFSIG13tecs}, but we don't make use of this here.}
The spatial robustness of trajectory $(\hotraj,\tstmp)$ starting at time $t$ w.r.t. formula $\formula$ is denoted as follows
\[\dle \formula \dri((\hotraj,\tstmp),t) = r \in \CoRe\]
Computing $r$ is done on the output trajectory without any reference to the system that generated it.

Time robustness measures by how much to shift the given trajectory \emph{in time}, to change its truth value w.r.t. $\formula$. 
Two time robustness values may be measured for each trajectory: the future robustness $\theta^+$ and the past robustness $\theta^-$, depending on whether the signal is shifted left (so future values are introduced) or right (so past values are introduced). In this work we explicitly denote time robustness by
\[\dle \formula \dri_\theta ((\hotraj,\tstmp),t) = \min \{\theta^-,\theta^+\} \in \CoRe \]
The spatial \cite{FainekosP09tcs} and temporal \cite{DonzeM_SignalTL10} robust semantics of MTL formulae are reviewed in the appendix.

Both types of robustness (spatial and temporal) satisfy the fundamental theorem that a negative robustness value indicates falsification, a positive value indicates satisfaction, and a value of 0 indicates that an infinitesimal change in the trajectory (in space or in time) will change its truth value.
Therefore, the search for a falsifying trajectory $\hotrajp$ can be re-cast as the problem of minimizing $\dle \formula \dri(\hotrajp,0)$ over $\hsSet_0 \times \inpSpace^{|N|}$.
To make this a finite-dimensional optimization, the input signals are parameterized with a finite number of parameters.
(This parametrization effectively limits the search space, and the global minimum returned by falsification is a minimum over this limited space. But the parametrization can typically be made as precise as desired, e.g. to within the approximation error of the minimization algorithm).
As our objective is to find falsifying trajectories, we stop the search as soon as it encounters a trajectory with negative robustness.

Now it is possible to create an example which displays a (graphically) convergent sequence of trajectories 
$ (\hotraj_{||,i},\tstmp_{||,i}) \rightarrow (\hotrajp,\tstmp_{||})$ 
such that $\dle \formula_{\tteps} \dri((\hotraj_{||,i},\tstmp_{||,i}),0)$ does not converge to $\dle \formula_{\tteps} \dri ((\hotrajp,\tstmp_{||}),0)$. This holds true for both spatial and temporal robustness.
So even if Model and Implementation are not conformant (for a given value of \te), local optimization algorithms can get trapped in local minima with positive robustness.
On the other hand, non-conformant trajectory pairs will necessarily have negative robustness, so that if a Model/Implementation pair is non-conformant, all global minima of the robustness are negative, and correspond to non-conformant pairs of trajectories. 
Thus we need to use global optimizers, like Simulated Annealing, Cross-Entropy~\cite{SankaranarayananF_CE12} or other methods supported by~\cite{AnnapureddyLFS11tacas}.

\subsection{Degree of conformance}
In addition to verifying whether two systems are $\tteps$-close for a given $\tteps$, we may find a smallest such pair with the order defined in Def.\ref{def:teorder}.
Recall now that $\formula_{\tteps}$ is monotonic in $\tteps$ (remark \ref{rem:monotonous}). 
The following theorem shows that the robustness values are also monotonic in the parameters $\tau, \varepsilon$. The proof is in Appendix~\ref{sec:proofMonotonic}.
\begin{thm}
\label{prop:monotonousrob}
Take two TSS $(\hotraj,\htstmp)$ and $(\hotraj',\htstmp')$, a test duration $T$, a number of jumps $J$, and a time $t \leq T$.
Consider the parallel concatenation  
\[(\hotrajp, \htstmpp) = ((\hotraj, \htstmp), (\hotraj', \htstmp'))\]
(i) Fix $\tau > 0$. If $0 < \varepsilon_1 \leq \varepsilon_2$, then 
\[\dle \formula_{(\tau,\varepsilon_1)} \dri ((\hotrajp, \htstmpp),t) \leq \dle \formula_{(\tau,\varepsilon_2)} \dri ((\hotrajp, \htstmpp),t)\]
(ii) Now fix $\varepsilon> 0$. If $0 < \tau_1 \leq \tau_2$, then
\[\dle \formula_{(\tau_1,\varepsilon)} \dri_\theta ((\hotrajp, \htstmpp),t) \leq \dle \formula_{(\tau_2,\varepsilon)} \dri_\theta ((\hotrajp, \htstmpp),t)\]
\end{thm}

Therefore, we can combine {\staliro} with a binary search over the values of $\tau$ and $\varepsilon$ to find a smallest pair such that $\formula_{\tteps}$ is satisfied. 
Because the order  on $\tteps$ pairs is only partial, binary search is applied to each component while fixing the other, thus exploring the Pareto-optimal front (e.g.~\cite{LegrielGCM_Pareto10}).
Algorithm \ref{alg:dichotomy} shows the binary search for the smallest $\varepsilon$ given a $\tau$. A search over $\tau$ can be done similarly with obvious modifications. 
The initial $\varepsilon_h$ can be found by using an initial binary search that doubles some $\varepsilon_0$ until $\dle \formula_{(\tau,\varepsilon)} \dri > 0$.

\begin{algorithm}
  \caption{Searching for a smallest $\varepsilon$ given $\tau$.}
  \label{alg:dichotomy}
  \begin{algorithmic}
    \Require Number of iterations $K$, parameter $\tau > 0$, low value $\varepsilon_l = 0$, high value $\varepsilon_h > 0$ such that $\dle \formula_{(\tau,\varepsilon_h)} \dri > 0$.		
		
	\For {$i=0$ to $K-1$}
		\State $\varepsilon = 0.5*(\varepsilon_h + \varepsilon_l)$
		\State Run {\staliro} to falsify $\formula_{(\tau,\varepsilon)}$.
		\If{ ($\dle \formula_{(\tau,\varepsilon)} \dri < 0$) }
		\State $\varepsilon_l = \varepsilon$ 
		\Else
		\State $\varepsilon_h = \varepsilon$;
		\EndIf		
	\EndFor
	
	\Return $[\varepsilon_l,\varepsilon_h]$

  \end{algorithmic}
\end{algorithm}

The value $(\bar \tau, \bar \varepsilon)$ returned by this procedure gives a \emph{quantitative measure of conformance between the two systems}, and allows the designer to make informed trade-offs between, say, output accuracy of the Implementation, and its timing characteristics.

\begin{rem}
\label{rem:expliciteps}
For a given $\tau$, the smallest $\varepsilon$ such that two trajectories $(\hotraj,\htstmp)$ and $(\hotraj',\htstmp')$ are $\tteps$-close can be calculated as
\begin{eqnarray}
\label{eq:expliciteps}
\varepsilon_M(\tau) &=& \min_{i \in N} \max_{|\htstmp'_k - \htstmp_i| < \tau} \|\hotraj_{i} - \hotraj'_{k}\| \\
\varepsilon_I(\tau) &=& \min_{i \in N} \max_{|\htstmp'_k - \htstmp_i| < \tau} \|\hotraj_{k} - \hotraj'_{i}\| \\
\varepsilon(\tau)   &=& \max \{\varepsilon_M(\tau),\varepsilon_I(\tau)\}
\end{eqnarray}
Similar definitions hold for the smallest $\tau$ given an $\varepsilon$.
We can minimize $\varepsilon(\tau)$ over the space of TSS to determine a smallest $(\tau,\varepsilon_*(\tau))$ such that the two systems are \tec.
The approach in Algorithm \ref{alg:dichotomy} has the advantage of working not just for \tec, but any other, application-dependent, notion of conformance.
\end{rem}

\section{Experiments}
\label{sec:experiments}

\begin{figure*}[t]
\centering
\begin{tabular}{cc}
\includegraphics[width=\columnwidth, height = 4.5cm]{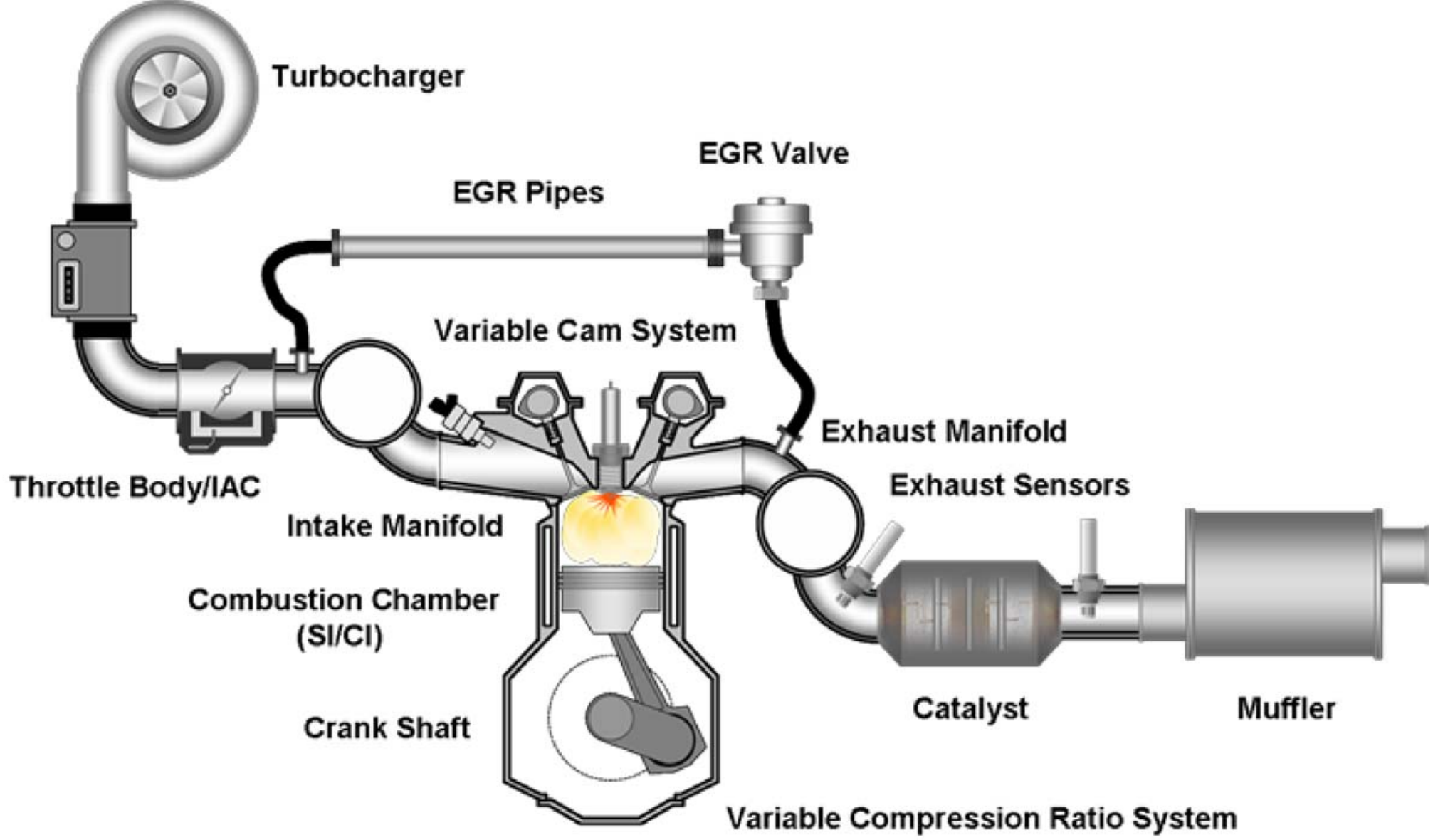} &
\includegraphics[width=\columnwidth, height = 4.5cm]{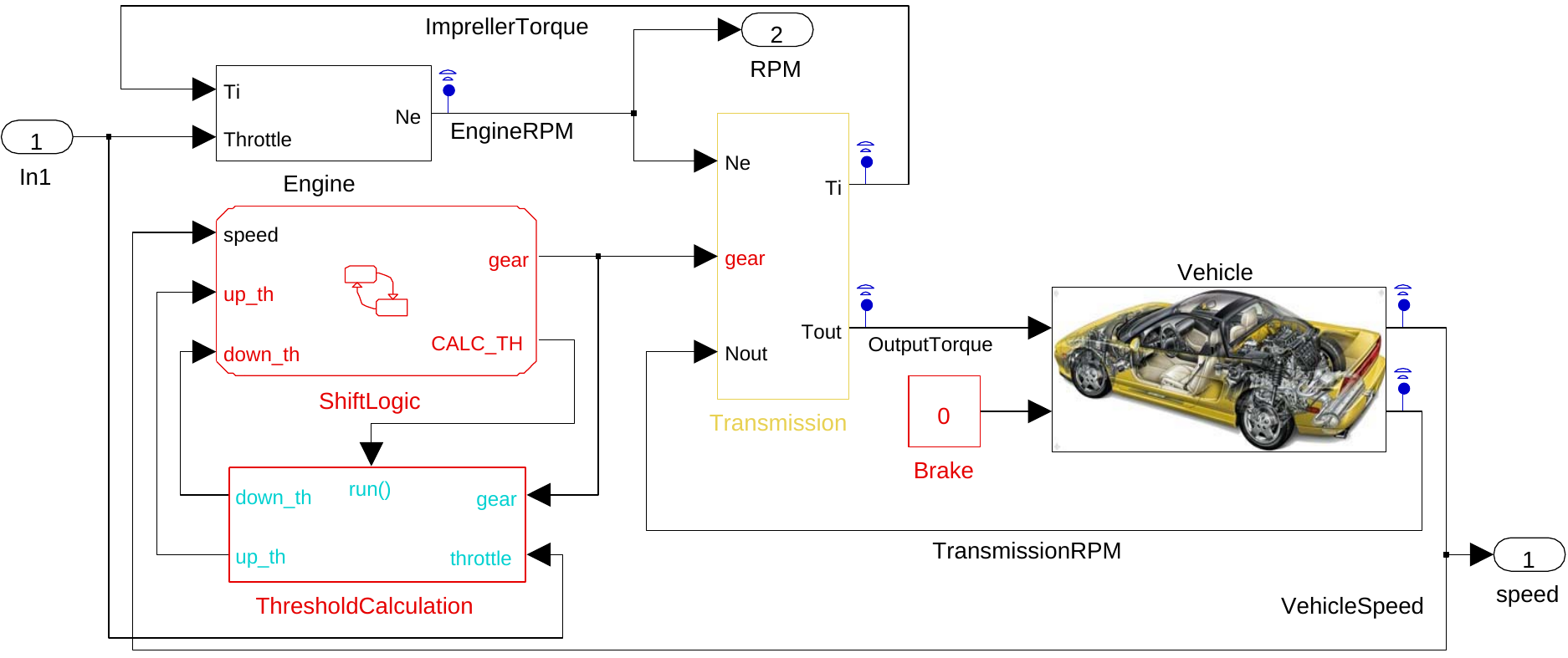} 
\end{tabular}
\vspace{-0.2cm}
\caption{Example \ref{ex:simuquest}. Left: SimuQuest Enginuity model components. Used with permission, \copyright SimuQuest\cite{Simuquest:Online}. Right: Automatic Transmission Model.}  
\label{fig:simuquestATEngine}
\vspace{-0.2cm}
\end{figure*}

We illustrate the proposed approach on three systems, including a commercial high fidelity engine model.
In all experiments, we didn't restrict the maximum number of jumps $J$ in a given trajectory; rather, the simulation ended only when simulation time reached $T$.
So below, we set $J$ equal to some appropriately large $J_M$. 

\begin{exmp}[Example \ref{ex:AFC} continued]
\label{ex:AFCtest}
We use the FC Model and Implementation from Example \ref{ex:AFC} to illustrate the application of Algorithm \ref{alg:dichotomy} to find the tightest values of $\tau$ and $\varepsilon$ such that \tec~is true.
Because the \teps~pairs are partially ordered, we are looking for the Pareto-optimal front.
We decided to fix $\tau$ at 0.01, and do a search over $\varepsilon$. To determine which value of $\varepsilon$ to start the search from, we computed the maximum relative error between the outputs of the LUTs and the outputs of the corresponding polynomials over a window of 85 seconds, using randomly generated inputs. 
The maximum relative error was 0.4091. Obviously, because the LUTs are deep in the system, we do not expect the same relative error at their outputs as that at the output of the entire system. 
However, this duplicates the typical procedure for deciding how many entries to have in an LUT: fewer levels consumes less memory and makes for a faster computation, but causes greater error. So the designer starts from a few entries and observes the output of the system. If the error in the oputput is not acceptable, entries are added to the LUT to provide a better approximation. And so on.

Figure \ref{fig:AFCtest} shows a close-up of the the output trajectories from System and Implementation. Note that, as shown in Fig.~\ref{fig:AFCFuel} for the Fuel output, the two trajectories don't simply diverge and maintain one distance from each other, but rather, they diverge for a period only to meet up again. This interplay between time difference and space difference is well-captured by \tec.

\begin{figure*}[t]
\centering
\begin{tabular}{c}
\includegraphics[width=16cm,height=5cm]{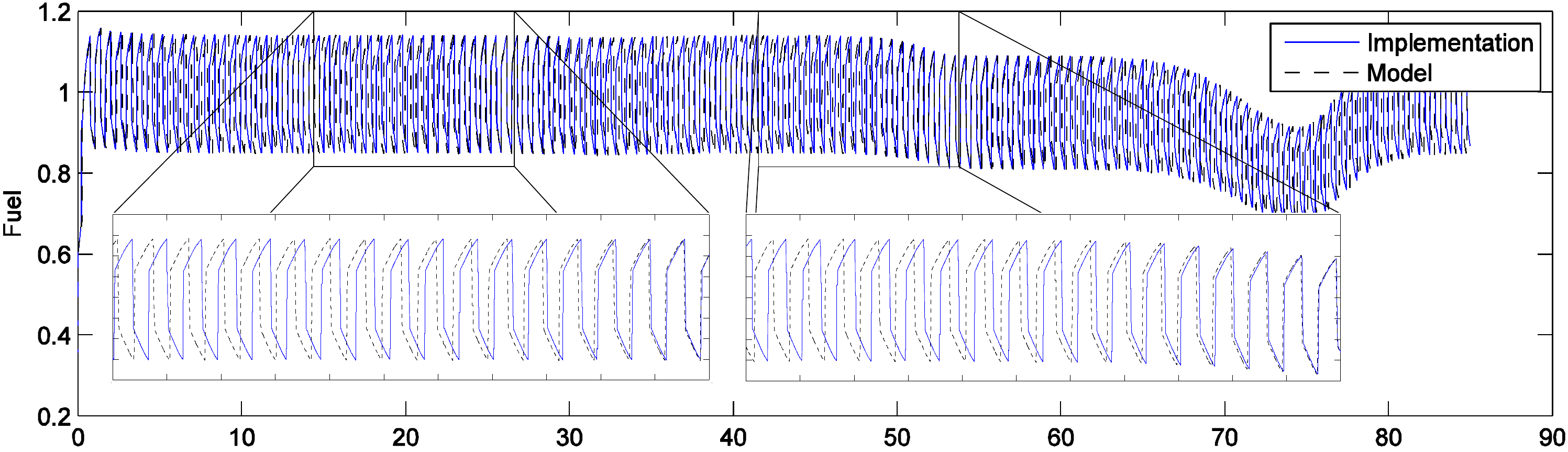}
\end{tabular}
\vspace{-0.4cm}
\caption{Example \ref{ex:AFCtest}. The Fuel output trajectories periodically separate from each other and converge again.}
\label{fig:AFCFuel}
\vspace{-0.2cm}
\end{figure*}

\begin{figure}[ht]
\vspace{-10pt} 
\begin{center}
\includegraphics[width=8cm]{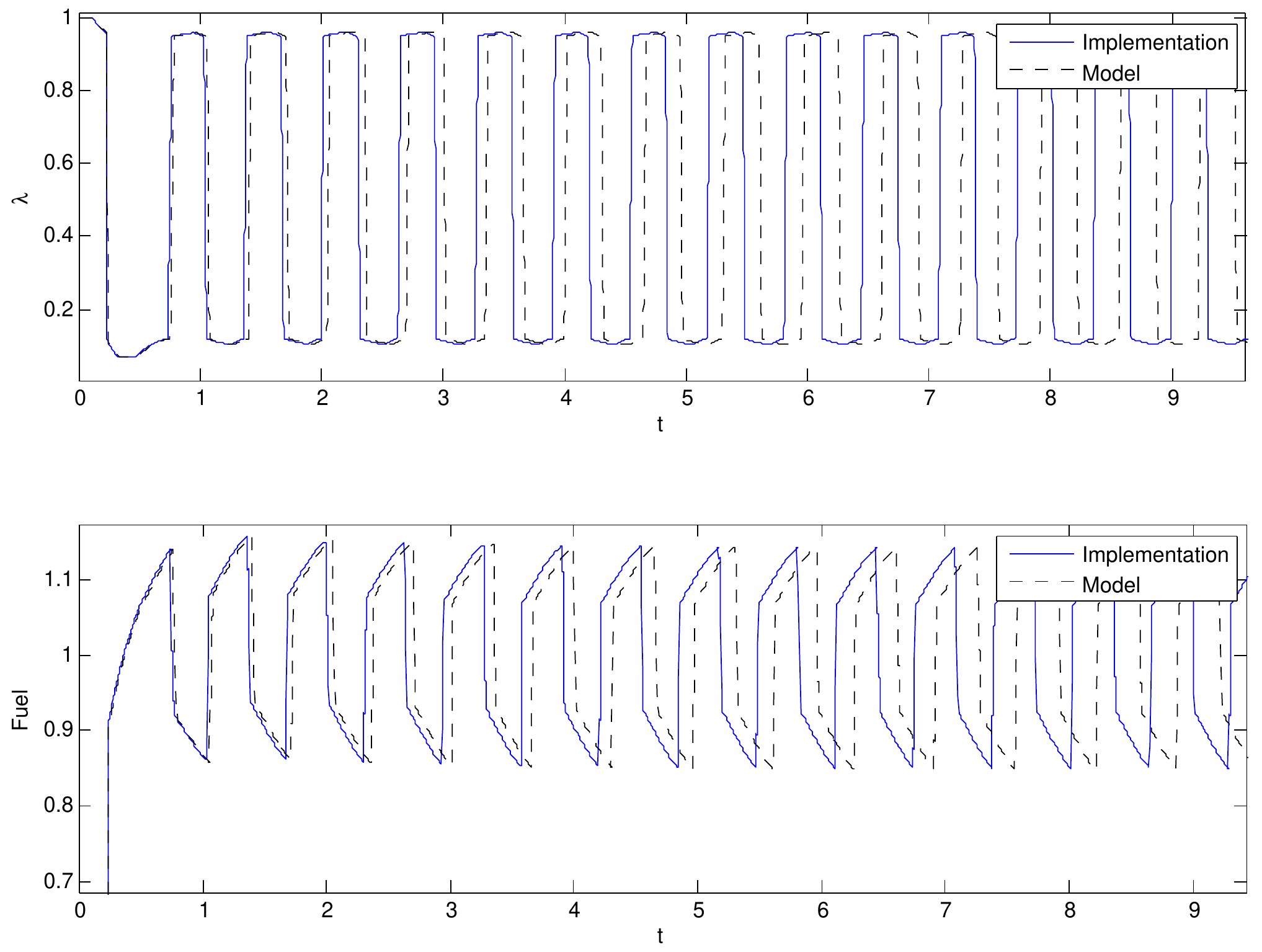}
\end{center}
\vspace{-15pt}
\caption{Example \ref{ex:AFCtest}. Close-up on the trajectories.}
\label{fig:AFCtest}
\vspace{-10pt}
\end{figure}
%

{\staliro} \cite{AnnapureddyLFS11tacas} was run at each iteration of the binary search to falsify $\formula_{(0.01,\varepsilon)}$. Algorithm \ref{alg:dichotomy} found an interval $\varepsilon \in [0.71752, 0.71832]$ over which the robustness varies between between $-0.0029$ and $0.027$. That is, The two systems are $(85,J_M,(0.01,\varepsilon))$-close with $\varepsilon \in [0.71752, 0.71832]$.
\end{exmp}


\begin{figure}
\includegraphics[width=\columnwidth]{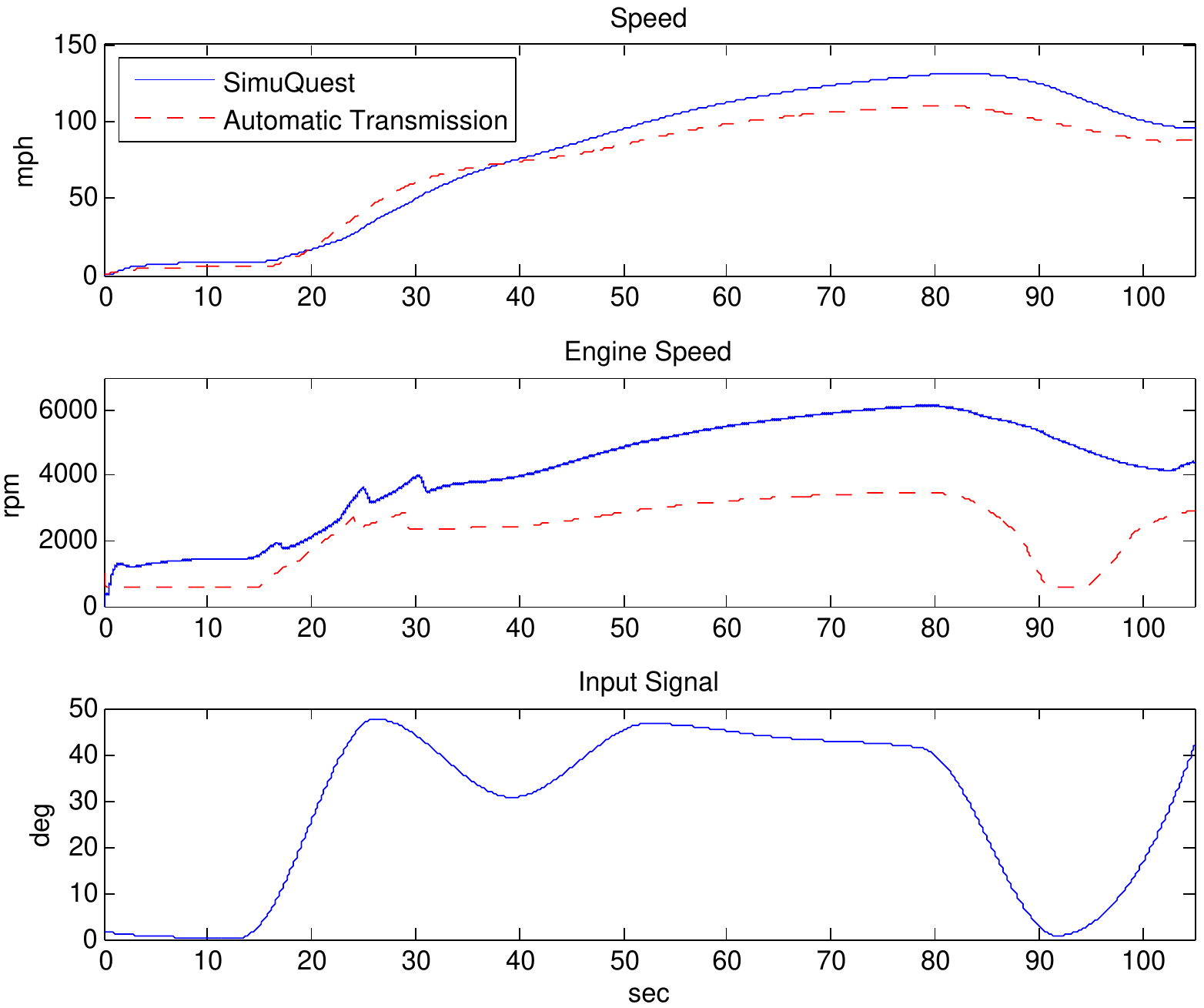}
\caption{Example \ref{ex:simuquest}. The output trajectories for the SimuQuest and Automatic Transmission Engine models that fail the $\tteps$-closeness specification.}
\label{fig:simuquest}
\end{figure}

\begin{exmp} [High fidelity engine model]
\label{ex:simuquest}
Our second experiment was performed on a Model and Implementation of an automatic transmission.
The transmission has one input (throttle angle), and two outputs: 
the speed of the engine $\omega$ (RPM) and the speed of the vehicle $v$ (MPH), i.e., $\hotraj = [\omega \; v]^T$.
Here too, the goal is to find a smallest $\tteps$ such that the two are systems are \teps-close.
The Model is a slightly modified version of the Automatic Transmission model provided by Mathworks as a Simulink demo\footnote{Available at: \url{http://www.mathworks.com/products/simulink/demos.html}}. The model is shown in Figure \ref{fig:simuquestATEngine} right.
It contains 69 blocks including 2 integrators, 3 look-up tables, 3 2D look-up tables and a Stateflow chart.
The Stateflow chart contains two concurrently executing Finite State Machines with 4 and 3 states, respectively.


The Implementation is the Enginuity model of a Port Fuel Injected spark ignition engine from Simuquest~\cite{Simuquest:Online} with 56 states and a large number of black box components. A overview of the components of the model is shown in Figure \ref{fig:simuquestATEngine} left.
It is significantly more complex than the Model, as it models the effects of combustion from first physics principles on a cylinder-by-cylinder basis, while also including regression models for particularly complex physical phenomena. 

The initial conditions $\stPt_0$ are the initial RPM and the initial vehicle speed, both of which must be 0. Therefore, $\stSet_0 = \{ [0 \; 0]^T \}$.
This means the output trajectories depend only on the input signal $\inpSig$.
The throttle at each point in time can take any value between 0 (fully closed) and 100 (fully open).
We remark that the system is deterministic, i.e., under the same input $\inpSig$, we will always observe the same output $\hotraj$.
Test duration is set to $T=104$secs.

In 31 iterations, binary search found an interval of [4.8833, 4.8834], over which the spatial robustness varies between -0.00013 and 0.03. 
Thus the Model and Implementation are $(104, J_M, (5e-4,\varepsilon))$-close with $\varepsilon \in [4.8833, 4.8834]$. In Figure \ref{fig:simuquest} we present two output trajectories that fail the $\tteps$-closeness specification given the same input sequence.
\end{exmp}
\begin{exmp}
\label{ex:nav0}
To illustrate the falsification of application-dependent notions, we choose $\formula_{PWC}$ given by \eqref{eq:npwc}, and apply it to the navigation benchmark Nav0 from \cite{AbbasATVA11_LinFalsification}. 
Nav0 is a 4D hybrid automaton with 16 modes. Its guard sets are categorized as either `horizontal' or `vertical'.
Fifteen implementations are generated by varying the continuous dynamics in each mode (resulting in Implementations Dyn$_1$-Dyn$_9$), and varying the horizontal guards (resulting in Implementations HG$_1$-HG$_3$) and vertical guards (resulting in Implementations VG$_1$-VG$_3$).
The variations are such that the difference between Nav0 and Dyn$_k$ is smaller than the difference between Nav0 and Dyn$_{k+1}$. Similarly, the difference between Nav0 and HG$_k$ is smaller than the difference between Nav0 and HG$_{k+1}$, and comparable to that between Nav0 and VG$_k$.

We ran {\staliro} to minimize $\dle \formula_{PWC} \dri_\theta$, the temporal robustness of $\formula_{PWC}$.
Simulated Annealing (SA) was used as optimizer. Since it is a stochastic algorithm, 
to collect statistics, we ran 20 runs of 500 tests each, and each test lasts for $T = 20$ seconds. $D$ was set to 0.5.
The results are presented in Table \ref{tab:nav0PWCSA}. 12 out of the 15 implementations were falsified, i.e. found to be non-conformant to the Model. 
Implementations HG$_{1-3}$ are robustly conformant to the Model, as their robustness was infinite: this means that modifying the horizontal guards within the amounts prescribed by HG$_3$ can not affect PWC conformance.
On the other hand, only one test was sufficient to falsify $\formula_{PWC}$ with the vertical guard modifications. This shows great sensitivity of the system to the vertical guard conditions.
This is useful design input, as it tells the designers that they can trade-off horizontal guard implementation accuracy for greater accuracy in implementing the vertical guards.
\begin{table*}[t]
	\centering
		\begin{tabular}{|c|c|c|c|c|}
		\hline
		\textbf{Implementations}    & Nb falsifying runs  &  Avg nb of tests            & Avg robustness      	& Avg falsification time \\
																&	(out of 20)  				& required for falsification  &      					        &   		 \\
		\hline
		$\mbox{Dyn}_1$              & 17                  &  181.47      								& $-\infty$   	& 153.12  \\
		\hline
		$\mbox{Dyn}_2$              & 13                  &  119.3      								& $ -\infty$   	& 98.08,\\
		\hline
		$\mbox{Dyn}_3$              & 18                  &  141.77      										& $ -\infty$   	& 117.71\\
		\hline
		$\mbox{Dyn}_4$              & 20                  &  41,45      										& $ -\infty$   			& 33.12\\
		\hline
		$\mbox{Dyn}_5$              & 20                  &  31.65       								& $ -\infty$   			& 24.82\\
		\hline
		$\mbox{Dyn}_6$              & 20                  &  27.55       									& $-\infty$   			& 21.71\\
		\hline
		$\mbox{Dyn}_7$              & 20                  &  11.6        										& $ -\infty$   			&  8.60\\
		\hline
		$\mbox{Dyn}_8$              & 20                  &  2.15        										&  -0.081   				& 1.59\\
		\hline
		$\mbox{Dyn}_9$              & 20                  &  1.15        												& $ -\infty$   		& 0.90\\
		\hline
		$\mbox{HG}_1$               & 0                  &  N/A               		& $+\infty$ 	& N/A\\
		\hline
		$\mbox{HG}_2$                           & 0                  &  N/A               		& $+\infty$ 	& N/A\\
		\hline
		$\mbox{HG}_3$                           & 0                  &  N/A               		& $+\infty$ 	& N/A\\
		\hline
		$\mbox{VG}_1$                           & 20                  &  1               & $-\infty$ 	&0.46\\
		\hline
		$\mbox{VG}_2$                           & 20                  &  1             & $-\infty$ 	& 0.47\\
		\hline
		$\mbox{VG}_3$                           & 20                  &  1               & $-\infty$ 	& 0.48\\		
		\hline
		\end{tabular}
	\caption{Results of minimizing $\dle \formula_{PWC} \dri_\theta$ using {\staliro}. For each Implementation, are given the number of runs that showed it to be non-conforming to the Model (second column), the average number of tests (or trajectories) needed before a falsifying trajectory is found (third column), the average robustness, and the average runtime until falsification. Average quantities are taken over the 20 runs.}
	\label{tab:nav0PWCSA}
\end{table*}
\end{exmp}

\section{Related Work}
\label{sec:related works}

Tretmans \cite{tretmans1999testing} defined Input-Output conformance (\textsc{ioco}) as requiring that the Implementation never produces an output that can not be produced by the specification, and it is never the case that the Implementation fails to produce an output when the specification requires one. Both Implementation and specification are modeled as (discrete) labeled transition systems. 
Van Osch \cite{Osch_IOCO06} later extended \textsc{ioco} to hybrid transition systems (HTS) by incorporating continuous-time inputs.
This hybrid \textsc{ioco} is not testable in practice because the state space and transition relations of an HTS are uncountable, and the test generation algorithm proposed in \cite{Osch_IOCO06} doesn't contain a mechanism for judiciously choosing tests from the infinite set of possible tests.

Later work \cite{WoehrleKT_conformance12} also extends \cite{tretmans1999testing} by treating the Implementation as a black box that generates timed traces, and representing the specification as a timed automaton.
The objective is to verify, for each trace generated by the Implementation, whether it satisfies the invariants of the specification automaton. 
As such, this conformance notion does does not address this paper's goal of verifying `similarity' between an Implementation and its Model, which is a more comprehensive problem.
The work by Brandl et al. \cite{brandl2010automated} utilizes (discrete) action systems \cite{back1990stepwise} to provide a discrete view of hybrid systems (a modeling formalism for CPS). Thus Tretmans' \textsc{ioco} can be applied to the now-discrete system. This method requires knowledge of the internal system structure, which we do not assume in our work.

In~\cite{AbateP_Abstractions11}, a distance between systems is also defined via a distance between trajectories. The closeness notion used there can be shown to be weaker than \tec, so that proving two systems to be \te-close implies they are close in the sense of~\cite{AbateP_Abstractions11}. In fact, \tec\; provides a continuum of closeness degrees between the two extremes presented in~\cite{AbateP_Abstractions11}.

\section{Conclusions}

In this paper, we have defined conformance between a Model and its Implementation as a degree of closeness between the outputs of the two systems.
This notion is quantifiable, thus allowing us to speak of degrees of conformance, giving a richer picture of the relation between the two systems.
It is also applicable to very general system models, which allows us to study the conformance of Models to complex Implementations.
This conformance was then expressed as an MTL formula, allowing us to use existing falsification tools to find non-conformant behavior of Model and Implementation, if it exists.

Because a CPS will usually have several operating modes with different dynamics, it will be interesting in future work to explicitly incorporate the mode switching into the MTL formulae.
Finally, a more complete theory of conformance should also account for different time domains between the Model's trajectories and the Implementation's trajectories.

\section{Acknowledgments} 
The work presented here benefited from the input of Raymond Turin, Founder and CTO at SimuQuest, who provided assistance in working with the SimuQuest Enginuity model. 

This work was partially funded under NSF awards CNS 1116136, CNS 1319560, IIP-0856090 and the NSF I/UCRC Center for Embedded Systems.

\appendix
\section{MTL Robustness}\nonumber
\label{sec:mtl}

In this section, we review the robust semantics of MTL formulas. 
Details on the theory and algorithms are available in our previous work~\cite{FainekosP09tcs,FainekosSUY12acc}. 

\begin{defn}[MTL Syntax] 
Let $AP$ be the set of atomic propositions and $\Ic$ be any non-empty interval of $\preals$.
The set $\mtl$ of all well-formed MTL formulas is inductively defined as
$ \varphi \; ::= \; \mathbf{T} \; | \; p \; | \; \neg \varphi \; | \; \varphi \vee \varphi \; | \; \varphi U_\Ic \varphi$,
where $p \in AP$ and $\mathbf{T}$ is {\em true}.
\end{defn}

We provide semantics that map an MTL formula $\varphi$ and an output trajectory $(\outtraj,\tstmp)$ of $\Sys$ to a value drawn from $\CoRe$. 
For an atomic proposition $p \in AP$, the semantics evaluated for $(\outtraj_i,\tstmp_i)$ consists of the distance between $\outtraj_i$ and the set $\Oc(p)$ labeling $p$.  
Intuitively, this distance represents how robustly the point $\outtraj_i$ lies within (or is outside) the set $\Oc(p)$.
If this distance is zero, then  the smallest perturbation of the point $\outtraj_i$  can affect the outcome of $\outtraj_i \in \Oc(p)$.  
We denote the spatial robust valuation of the formula $\varphi$ over the trajectory $(\outtraj,\tstmp)$ at time $t$ by $\dle \varphi, \Oc \dri((\outtraj,\tstmp),t)$. 
Here $t$ is such that $t = \tstmp_i$ for some $i \in N$.
The solution $\outtraj$ always starts from time $\tstmp_1 =0$.
Formally, $\dle \cdot, \cdot \dri : (\mtl \times \Omap) \rightarrow (\Sig \times [0,T] \rightarrow \CoRe)$.

\begin{defn} [Robust Semantics] 
Let $\pi = (\outtraj,\tstmp)$ be a real-TSS output of (\ref{eq:sys}) and $\Oc \in \Omap$, and let $\Ic$ be a non-empty interval on the real line. Then the robust semantics of any formula $\varphi \in \mtl$ with respect to $\pi$ is defined as:
\begin{align*} 
\allowdisplaybreaks
\dle \mathbf{T}, \Oc \dri (\pi,t) := &   +\infty
\displaybreak[2] \\
\dle p, \Oc \dri (\pi,t) := &  \mathbf{Dist} (\pi(t),\Oc(p)) 
\displaybreak[2] \\
\dle \neg \varphi_1, \Oc \dri  (\pi,t)  :=  & - \dle \varphi_1, \Oc \dri (\pi,t) 
\displaybreak[2] \\
\dle \varphi_1 \vee \varphi_2, \Oc \dri (\pi,t) := &  \dle \varphi_1, \Oc \dri  (\pi,t) \sqcup \dle \varphi_2, \Oc \dri  (\pi,t)
\displaybreak[2] \\
\dle \varphi_1 U_\Ic \varphi_2, \Oc \dri  (\pi,t) := &  \bigsqcup_{t' \in (t+_{[0,T]}\Ic)} ( \dle \varphi_2, \Oc \dri (\pi,t') \sqcap\\
&  \sqcap_{t \leq t'' <t'} \dle \varphi_1, \Oc \dri (\pi,t'') 
\end{align*}
where $\mathbf{Dist} (z,S)$ is the signed distance of $z \in X$ from a set $S \subseteq X$
\[ \mathbf{Dist}(z,S) := \left\{ 
\begin{array}{ll}
- \inf\{\| z - z' \| \; | \; z' \in S \} & \mbox{ if } z \not \in S \\
\inf\{\| z - z' \| \; | \; z' \in X \backslash S \} & \mbox{ if } z \in S \\
\end{array} \right. \]
where $t+_{[0,T]}\Ic = \{ t'' \in [0,T] \; | \; \exists t' \in \Ic \, .\, t'' = t + t' \}$,
$\sqcup$ and $\sqcap$ stand for the supremum and infimum, respectively,
and $\sup \emptyset := -\infty$ and $\inf \emptyset := +\infty$.
The semantics of the other operators can be defined using the above basic operators.
E.g., $\Diamond_\Ic \phi \equiv {\mathbf T} U_\Ic \phi$ and $\Box_\Ic \phi \equiv \neg \Diamond_\Ic \neg \phi$.
\label{def:mitlrob}
\end{defn}

It can be shown \cite{FainekosP09tcs} that if the signal satisfies the property, then its robustness is non-negative, and if the signal does not satisfy the property, then its robustness is non-positive.  

The time robust semantics differ from the above only in the definition of the base case. 
Take $t$ such that $\tstmp_i = t$ for some $i\in N$.
If we let $p[t]$ denote the truth value of $ \pi_i \models \formula$, then 
\begin{gather*}
\theta^-(p,\pi,t) \defeq p[t]\cdot \max \{d \geq 0 \st d = \tstmp_i - \tstmp_k \text{ and } \\
\forall k \leq q \leq i, p[\tstmp_q] = p[t]\} \\
\theta^+(p,\pi,t) \defeq p[t]\cdot \max \{d \geq 0 \st d = \tstmp_k - \tstmp_i \text{ and } \\
 \forall k \geq q \geq i, p[\tstmp_q] = p[t]\} \\
\dle p \dri_\theta(\pi,t) = \min \{\theta^-,\theta^+\}
\end{gather*}
The rest of the equations above follows through unchanged.

\section{Proof of Theorem 4.1}
\label{sec:proofMonotonic}

\newcommand \dy {d}
\newcommand \ds {\tstmp}
We start by proving the result for real-TSS. The extension to hybrid-TSS will then follow immediately.
So start by considering that $(\hotraj,\tstmp)$ and $(\hotraj',\tstmp')$ are real-TSS, and for convenience, we will use $\pi$ to denote their parallel concatenation $(\hotrajp, \htstmpp)$. 
Recall \eqref{eq:p1},\eqref{eq:p2}, and the robust semantics of MTL from Appendix~\ref{sec:mtl}.

(i) Define $\dy_k =  \|\hotraj_{M,i} - (\Shift_k \hotraj_I)_i\|$, 
and the atomic proposition 
$a^k_\varepsilon \defeq \dy_k < \varepsilon$.
Equation \eqref{eq:p1} can be written as 
\[p_1(\tau,\varepsilon) = \lor_{k=-n(\tau)}^{n(\tau)} a^k_\varepsilon \]
So it holds that 
\[\dle a^k_\varepsilon \dri(\pi,t) = \mathbf{Dist}(\dy_k, (-\varepsilon,+\varepsilon)) = \varepsilon - \dy_k\]
Thus with $\varepsilon_1 \leq \varepsilon_2$, $\dle a^k_{\varepsilon_1} \dri(\pi,t) \leq \dle a^k_{\varepsilon_2} \dri(\pi,t)$.
By the robust semantics, 
\begin{gather*}
\dle p_1(\tau,\varepsilon_1) \dri(\pi,t) = \max_k \{\dle a^k_{\varepsilon_1} \dri(\pi,t) \} \\
\leq \max_k \{\dle a^k_{\varepsilon_2} \dri(\pi,t) \} = \dle p_1(\tau, \varepsilon_2) \dri(\pi,t)
\end{gather*}
Similarly, we can show $\dle p_2(\tau,\varepsilon_1) \dri(\pi,t) \leq \dle p_2(\tau, \varepsilon_2) \dri(\pi,t)$.
Thus 
\begin{gather*}
\dle \formula_{(\tau,\varepsilon_1)} \dri = \min \{\dle p_1(\tau,\varepsilon_1) \dri(\pi,t), \dle p_2(\tau,\varepsilon_1) \dri(\pi,t)\} \\
\leq \min \{\dle p_1(\tau,\varepsilon_2) \dri(\pi,t), \dle p_2(\tau,\varepsilon_2) \dri(\pi,t)\} = \dle \formula_{(\tau,\varepsilon_2)} \dri
\end{gather*}

(ii) The time parameter $\tau$ controls the numbers $n(\tau)$ and $m(\tau)$, i.e. the number of shifted versions of the signals that must be created. See \eqref{eq:p1},\eqref{eq:p2}.
An increase in $\tau$ can only lead to an increase in the number of shifted versions, i.e. $m(\tau)$ and $n(\tau)$ are both non-decreasing in $\tau$. 
Therefore $m(\tau_1) \leq m(\tau_2)$ and $n(\tau_1) \leq n(\tau_2)$.
Then by the robust semantics, 
\begin{eqnarray*}
\dle p_1(\tau_1, \varepsilon) \dri_\theta(\pi,t) & = & \max \{\dle a^k_{\varepsilon} \dri_\theta(\pi,t) \st -n(\tau_1) \leq k  \leq n(\tau_1)\} \\
& \leq & \max \{\dle a^k_{\varepsilon} \dri_\theta(\pi,t) \st -n(\tau_2) \leq k  \leq n(\tau_2)\} \\
& = & \dle p_1(\tau_2, \varepsilon) \dri_\theta(\pi,t)
\end{eqnarray*} 
since the maximization for $\tau_2$ is happening over a larger set.
Similarly, $\dle p_2(\tau_1, \varepsilon) \dri_\theta(\pi,t) \leq \dle p_2(\tau_2, \varepsilon) \dri_\theta(\pi,t)$.
Finally, 
$\dle \formula_{(\tau_1,\varepsilon)}\dri_\theta(\pi,t) \leq \dle \formula_{(\tau_2,\varepsilon)} \dri_\theta(\pi,t)$.

The extension to hybrid-TSS is straighforward: by \eqref{eq:phitauepsHtss}, 
\begin{gather*}
\dle \Phi_{(\tau,\varepsilon_1)}\dri(\pi,t) = \min_i \dle \formula^i_{(\tau,\varepsilon_1)}\dri(\pi,t) \\
               \leq \min_i \dle \formula^i_{(\tau,\varepsilon_2)}\dri(\pi,t) = \dle \Phi_{(\tau,\varepsilon_1)}\dri(\pi,t) 
\end{gather*}
and similarly
\[
\dle \Phi_{(\tau_1,\varepsilon)}\dri_\theta(\pi,t) \leq \dle \Phi_{(\tau_2,\varepsilon)} \dri_\theta(\pi,t)
\]

%
\bibliographystyle{abbrv}
\bibliography{fainekos_bibrefs}

\begin{thebibliography}{10}

\bibitem{AbateP_Abstractions11}
A.~Abate and M.~Prandini.
\newblock Approximate abstractions of stochastic systems: A randomized method.
\newblock In {\em Decision and Control and European Control Conference
  (CDC-ECC), 2011 50th IEEE Conference on}, pages 4861--4866, 2011.

\bibitem{AbbasATVA11_LinFalsification}
H.~Abbas and G.~Fainekos.
\newblock Linear hybrid system falsification through local search.
\newblock In {\em Automated Technology for Verification and Analysis}, volume
  6996 of {\em LNCS}, pages 503--510. Springer, 2011.

\bibitem{AbbasF_HybridSA12}
H.~Abbas and G.~Fainekos.
\newblock Convergence proofs for simulated annealing falsification of safety
  properties.
\newblock In {\em Proc. of 50th Annual Allerton Conference on Communication,
  Control, and Computing}. IEEE Press, 2012.

\bibitem{AbbasF_NonlinearDescent13}
H.~Abbas and G.~Fainekos.
\newblock Computing descent direction of mtl robustness for non-linear systems.
\newblock In {\em American Control Conference (ACC)}, pages 4411--4416, 2013.

\bibitem{AbbasFSIG13tecs}
H.~Abbas, G.~E. Fainekos, S.~Sankaranarayanan, F.~Ivancic, and A.~Gupta.
\newblock Probabilistic temporal logic falsification of cyber-physical systems.
\newblock {\em ACM Transactions on Embedded Computing Systems}, 12(s2), May
  2013.

\bibitem{AnnapureddyLFS11tacas}
Y.~S.~R. Annapureddy, C.~Liu, G.~E. Fainekos, and S.~Sankaranarayanan.
\newblock S-taliro: A tool for temporal logic falsification for hybrid systems.
\newblock In {\em Tools and algorithms for the construction and analysis of
  systems}, volume 6605 of {\em LNCS}, pages 254--257. Springer, 2011.

\bibitem{Antoulas2000}
A.~C. Antoulas, D.~C. Sorensen, and S.~Gugercin.
\newblock A survey of model reduction methods for large-scale systems.
\newblock {\em Contemporary Mathematics}, 280:193--219, 2000.

\bibitem{bugscope}
Atrenta.
\newblock Bugscope$\;^{TM}$.
\newblock [Online at: http://www.atrenta.com/solutions/bugscope.htm5].

\bibitem{back1990stepwise}
R.~Back and K.~Sere.
\newblock Stepwise refinement of parallel algorithms.
\newblock {\em Science of Computer Programming}, 13(2):133--180, 1990.

\bibitem{Bako_ModeEstimation13}
L.~Bako, V.~L. Le, F.~Lauer, and G.~Bloch.
\newblock Identification of {MIMO} switched state-space models.
\newblock In {\em American Control Conference (ACC), 2013}, pages 71--76, 2013.

\bibitem{brandl2010automated}
H.~Brandl, M.~Weiglhofer, and B.~K. Aichernig.
\newblock Automated conformance verification of hybrid systems.
\newblock In {\em Quality Software (QSIC), 10th International Conference on},
  pages 3--12. IEEE, 2010.

\bibitem{Caspi2002}
P.~Caspi and A.~Benveniste.
\newblock Toward an approximation theory for computerized control.
\newblock In {\em Embedded Software}, volume 2491 of {\em LNCS}, pages
  294--304. Springer, 2002.

\bibitem{DonzeM_SignalTL10}
A.~Donze and O.~Maler.
\newblock Robust satisfaction of temporal logic over real-valued signals.
\newblock In {\em Formal Modeling and Analysis of Timed Systems}, volume 6246
  of {\em LNCS}, pages 92--106. Springer Berlin Heidelberg, 2010.

\bibitem{Fainekos_staliro}
G.~Fainekos.
\newblock {\staliro}.
\newblock [Online at: https://sites.google.com/a/asu.edu/s-taliro/s-taliro].

\bibitem{FainekosP09tcs}
G.~Fainekos and G.~Pappas.
\newblock Robustness of temporal logic specifications for continuous-time
  signals.
\newblock {\em Theoretical Computer Science}, 410(42):4262--4291, September
  2009.

\bibitem{FainekosSUY12acc}
G.~Fainekos, S.~Sankaranarayanan, K.~Ueda, and H.~Yazarel.
\newblock Verification of automotive control applications using s-taliro.
\newblock In {\em Proceedings of the American Control Conference}, 2012.

\bibitem{FrehseCAV11}
G.~Frehse, C.~L. Guernic, A.~Donze, S.~Cotton, R.~Ray, O.~Lebeltel, R.~Ripado,
  A.~Girard, T.~Dang, and O.~Maler.
\newblock {SpaceEx}: Scalable verification of hybrid systems.
\newblock In {\em Proceedings of the 23d CAV}, 2011.

\bibitem{Girard08}
A.~Girard, A.~Julius, and G.~Pappas.
\newblock Approximate simulation relations for hybrid systems.
\newblock {\em Discrete Event Dynamic Systems}, 18(2):163--179, 2008.

\bibitem{GoebelT06_SolnsHybInclusions}
R.~Goebel and A.~Teel.
\newblock Solutions to hybrid inclusions via set and graphical convergence with
  stability theory applications.
\newblock {\em Automatica}, 42(4):573 -- 587, 2006.

\bibitem{Guz2010}
L.~Guzzella and C.~Onder.
\newblock {\em Introduction to Modeling and Control of Internal Combustion
  Engine Systems}.
\newblock Springer-Verlag, 2nd edition, 2010.

\bibitem{HenzingerMP_FORMATS05}
T.~A. Henzinger, R.~Majumdar, and V.~S. Prabhu.
\newblock Quantifying similarities between timed systems.
\newblock In {\em FORMATS}, volume 3829 of {\em LNCS}, pages 226--241.
  Springer, 2005.

\bibitem{HuangM2012}
Z.~Huang and S.~Mitra.
\newblock Computing bounded reach sets from sampled simulation traces.
\newblock In {\em The 15th International Conference on Hybrid Systems:
  Computation and Control}. {ACM}, 2012.

\bibitem{peet_verifPlan04}
P.~James.
\newblock {\em Verification Plans: The Five-Day Verification Strategy for
  Modern Hardware Verification Languages}.
\newblock Kluwer Academic Publishers, 2004.

\bibitem{Jin14}
X.~Jin, J.~Kapinski, J.~V. Deshmukh, K.~Ueda, and K.~Butts.
\newblock Fuel control system verification benchmark problems.
\newblock In {\em Submitted to: Hybrid Systems: Computation and Control}, 2014.

\bibitem{Lygeros99_Zeno}
K.~H. Johansson, J.~Lygeros, S.~Sastry, and M.~Egerstedt.
\newblock Simulation of hybrid zeno automata.
\newblock In {\em Conference on Decision and Control}, volume~4, pages
  3538--3543, December 1999.

\bibitem{Koymans90}
R.~Koymans.
\newblock Specifying real-time properties with metric temporal logic.
\newblock {\em Real-Time Systems}, 2(4):255--299, 1990.

\bibitem{Lecchini10_ConvRates}
A.~Lecchini-Visintini, J.~Lygeros, and J.~Maciejowski.
\newblock Stochastic optimization on continuous domains with finite-time
  guarantees by markov chain monte carlo methods.
\newblock {\em Automatic Control, IEEE Transactions on}, 55(12):2858 --2863,
  dec. 2010.

\bibitem{LeeS11book}
E.~A. Lee and S.~A. Seshia.
\newblock {\em Introduction to Embedded Systems: A Cyber-Physical Systems
  Approach}.
\newblock Online \url{http://leeseshia.org/}, 2011.

\bibitem{LegrielGCM_Pareto10}
J.~Legriel, C.~Guernic, S.~Cotton, and O.~Maler.
\newblock Approximating the pareto front of multi-criteria optimization
  problems.
\newblock In J.~Esparza and R.~Majumdar, editors, {\em Tools and Algorithms for
  the Construction and Analysis of Systems}, volume 6015 of {\em Lecture Notes
  in Computer Science}, pages 69--83. Springer Berlin Heidelberg, 2010.

\bibitem{LygerosJSZS03tac}
J.~Lygeros, K.~H. Johansson, S.~N. Simic, J.~Zhang, and S.~Sastry.
\newblock Dynamical properties of hybrid automata.
\newblock {\em IEEE Transactions on Automatic Control}, 48:2--17, 2003.

\bibitem{MalerNickovic04}
O.~Maler and D.~Nickovic.
\newblock Monitoring temporal properties of continuous signals.
\newblock In {\em Proceedings of FORMATS-FTRTFT}, volume 3253 of {\em LNCS},
  pages 152--166, 2004.

\bibitem{MazziEtAl08cdc}
E.~Mazzi, A.-S. Vincentelli, A.~Balluchi, and A.~Bicchi.
\newblock Hybrid system reduction.
\newblock In {\em 47th IEEE Conference on Decision and Control}, pages
  227--232, 2008.

\bibitem{Osch_IOCO06}
M.~Osch.
\newblock Hybrid input-output conformance and test generation.
\newblock In {\em Formal Approaches to Software Testing and Runtime
  Verification}, volume 4262 of {\em LNCS}, pages 70--84. Springer Berlin
  Heidelberg, 2006.

\bibitem{PlatzerQ08ijcar}
A.~Platzer and J.-D. Quesel.
\newblock {KeYmaera}: A hybrid theorem prover for hybrid systems.
\newblock In {\em International Joint Conference on Automated Reasoning},
  volume 5195 of {\em LNCS}, pages 171--178. Springer, 2008.

\bibitem{RoyTM11hscc}
P.~Roy, P.~Tabuada, and R.~Majumdar.
\newblock Pessoa 2.0: a controller synthesis tool for cyber-physical systems.
\newblock In {\em Proceedings of the 14th international conference on Hybrid
  systems: computation and control}, pages 315--316, New York, NY, USA, 2011.
  ACM.

\bibitem{Sanfelice11_interconnections}
R.~G. Sanfelice.
\newblock Interconnections of hybrid systems: Some challenges and recent
  results.
\newblock {\em Journal of Nonlinear Systems and Applications}, 2(1-2):111--121,
  2011.

\bibitem{SanfeliceT10automatica}
R.~G. Sanfelice and A.~R. Teel.
\newblock Dynamical properties of hybrid systems simulators.
\newblock {\em Automatica}, 46(2):239--248, 2010.

\bibitem{SankaranarayananF_CE12}
S.~Sankaranarayanan and G.~Fainekos.
\newblock Falsification of temporal properties of hybrid systems using the
  cross-entropy method.
\newblock In {\em ACM International Conference on Hybrid Systems: Computation
  and Control}, 2012.

\bibitem{Simuquest:Online}
Simuquest.
\newblock {Enginuity}.
\newblock \url{http://www.simuquest.com/products/enginuity}.
\newblock Accessed: 2013-10-04.

\bibitem{Tabuada2009}
P.~Tabuada.
\newblock {\em Verification and Control of Hybrid Systems: A Symbolic
  Approach}.
\newblock Springer, 2009.

\bibitem{Tiwari12cav}
A.~Tiwari.
\newblock {HybridSAL} relational abstracter.
\newblock In {\em Computer Aided Verification}, volume 7358 of {\em LNCS},
  pages 725--731. Springer, 2012.

\bibitem{tretmans1999testing}
J.~Tretmans.
\newblock Testing concurrent systems: A formal approach.
\newblock In {\em CONCUR 1999 Concurrency Theory}, pages 46--65. Springer,
  1999.

\bibitem{WoehrleKT_conformance12}
M.~Woehrle, K.~Lampka, and L.~Thiele.
\newblock Conformance testing for cyber-physical systems.
\newblock {\em ACM Trans. Embed. Comput. Syst.}, 11(4):84:1--84:23, Jan. 2013.

\bibitem{WongpiromsarnEtAl2011hscc}
T.~Wongpiromsarn, U.~Topcu, N.~Ozay, H.~Xu, and R.~M. Murray.
\newblock Tulip: a software toolbox for receding horizon temporal logic
  planning.
\newblock In {\em Proceedings of the 14th international conference on Hybrid
  systems: computation and control}, pages 313--314. ACM, 2011.

\end{thebibliography}
%
%


\end{document}